\documentclass[]{jfm}
\usepackage{graphicx}
\usepackage{siunitx}
\usepackage{newtxtext}
\usepackage{newtxmath}
\usepackage{natbib}
\usepackage{hyperref}
\hypersetup{
    colorlinks = true,
    urlcolor   = blue,
    citecolor  = black,
}

\usepackage{xcolor}
\usepackage{epstopdf, epsfig}
\usepackage{caption}
\usepackage{subcaption}
\usepackage{amsmath}
\usepackage{amssymb}
\usepackage{tikz}
\usetikzlibrary{shapes}
\definecolor{blue}{RGB}{0,112,192}
\definecolor{lightblue}{RGB}{0,176,240}
\definecolor{green}{RGB}{0,176,80}
\definecolor{yellow}{RGB}{255,255,0}
\definecolor{orange}{RGB}{255,192,0}
\definecolor{red}{RGB}{255,0,0}
\definecolor{darkred}{RGB}{118,0,0}
\definecolor{purple}{RGB}{208,0,154}

\newcommand{\RomanNumeralCaps}[1]
\linenumbers

\shorttitle{Visco-collisional rheology of dense granular-fluid flows}
\shortauthor{T. Man, H. E. Huppert, Q. Feng, K. M. Hill}
\title{Particle-scale studies elucidating visco-collisional rheology in granular-fluid flows}

\author{Teng Man\aff{1,2}\corresp{\email{manteng0520@zjut.edu.cn}},
  Herbert E. Huppert\aff{3},
  Qingfeng Feng\aff{2,4}
 \and Kimberly M. Hill\aff{2}
 \corresp{\email{kmhill@umn.edu}}
}

\affiliation{\aff{1}College of Civil Engineering, Zhejiang University of Technology, 288 Liuhe Rd, Hangzhou, Zhejiang 310023, China
\aff{2}Department of Civil, Environmental, and Geo- Engineering, University of Minnesota, 500 Pillsbury Drive SE, Minneapolis, Minnesota 55455, United States
\aff{3}Institute of Theoretical Geophysics, King's College, University of Cambridge, King's Parade, Cambridge CB2 1ST, United Kingdom
\aff{4}State Key Laboratory of Hydroscience and Engineering, Department of Hydraulic Engineering, Tsinghua University, Beijing100084, China
}
\begin{document}

\maketitle

\begin{abstract}
We study the particle-scale dynamics that give rise to bulk flow behaviours of highly concentrated particle-fluid mixtures using discrete element method (DEM) simulations. We utilize boundary conditions of a stress-controlled shear cell and vary material properties and applied stresses systematically.  We find the bulk rheology transitions smoothly among what might be called viscous, collisional, and visco-collisional behaviours based on average shear rate dependence as well as dominance of local interaction.  Using specific measures of particle-scale dynamics, we find that the transitions in system rheologies coincide with statistically significant changes in the ``fabric'' (i.e., strong force network and coordination numbers), the ``granular temperature'' (i.e., fluctuation energy) and relative importance of fluid and interparticle contacts.  Armed with these measures and previous theoretical work, we provide a foundation for understanding the particle-scale physics that gives rise to meso-scale rheologies and transitions from one to the next.  We determine physics-based formulations to better represent dynamic behaviors for a wide range of material properties and loading conditions. Our study broadens the applicability of rheological models to natural and engineered systems by providing a foundation for a more general constitutive model.
\end{abstract}


\section{Introduction}
\label{sec:intro}
Sheared particle-fluid flows are ubiquitous in natural and man-made systems, from muddy geophysical flows to hot mixed asphalt concrete \citep{man2022two,ZHANG2025106865}. Of these, those that flow at high solid concentrations, near the critical state [e.g., \citep{xing2024origin}] are elusive. The complexity of modelling these flows lies in part in mesoscale ephemeral structures that influence internal resistance to flow \citep{Man2025jfm}.  In turn, accurate representations of these structures are complicated by the indeterminacy of multiply-redundant (or occurring) dissipation mechanisms associated with contact deformations, frictional interactions, and viscous drag. Hence, phenomenologically determined macroscopic rheological relationships are not easily linked to the physics determined at the particle scale.  Computational models such as discrete element method (DEM) simulations can assist with parcing the disparate contributions, though, admittedly, they are limited to the accuracy of the manner with which the local interactions are represented.  In this paper, we consider two types of models that have compared well with experimental systems to propose a framework for linking the microscopic physics to the macroscopic phenomenology.  

Arguably, \citet{einstein1905motion,einstein1911neue} presented the first physics-based expression for a rheology of neutrally buoyant particles in a Newtonian fluid of viscosity $\eta_f$. Specifically, he showed the relationship between shear stress $\tau$ and strain rate $\dot {\gamma}$ to be directly analogous to that of a Newtonian fluid: $\tau=\eta\dot{\gamma}$, where, for a sufficiently low solid volume fraction $\phi$ ($\phi<0.02$) the effective viscosity of the suspension $\eta \approx \eta_{\rm{fl}}\times (2.5+\phi)$, where $\eta_{\rm{fl}}$ is the dynamic viscosity of fluid.  Nearly a half century later, \cite{bagnold1954experiments} demonstrated the relevance of normal stress associated with particle contacts, $\sigma_p$. More here on dependence of stresses on shear rates. 
Theories developed for building physics-based rheological models of their flows are primarily limited to cases with steady-state conditions and other relatively simple boundary conditions.

In the past two decades, significant progress has been made in capturing the rheology of these systems -- first, dry particle flows  \citep{pouliquen2002,midi2004,jop2006,pouliquen2006}, then viscosity-dominated fluid-particle flows \citep{cassar2005,boyer2011,trulsson2012}, and more recently, particle-fluid flows under a wider range of properties and boundary conditions \citep{Tapia2022, Ge2023unifying}.  Much of the foundation for these has involved ratios of local time scales: (1) inverse shear rate $1/ \dot \gamma$ and (2) interactive (inertial and/or viscous) timescales.
\begin{subequations}\label{eq:inertNum}
\begin{align}
I_c=\frac{\dot\gamma}{\sqrt{\sigma/(\rho_{p} d^2)}}\ \label{eq:1a}\\ 
\textrm{and}\ \ I_v=\frac{\dot\gamma}{ \sigma/\eta_f}\ , \label{eq:1b}
\end{align}
\end{subequations}
\noindent
typically referred to as \textit{inertial} and \textit{viscous} numbers, respectively. In a 2-d sheared system, $\sigma$ is the local stress measured normal to the average displacement direction; $d$ and $\rho_p$ are the particle size and density, respectively, and $\eta_f$ is the dynamic fluid viscosity.  The local steady-state rheology is expressed primarily in terms of two time-averaged local quantities: (1) a ratio of shear stress, $\mu_{\rm{eff}}=\tau / \sigma$, and (2) the solid volume fraction, $\phi$.  Table \ref{table:1} provides some widely validated functional forms of $\mu$ and $\phi$ found by empirical fits to data for dry flows (row 1) and viscosity-dominated suspensions (row 2).  Many other studies have supported these models (with similar fit coefficients) including (but are not limited to) \citep{pouliquen2002,midi2004,cassar2005,pouliquen2006,jop2006,boyer2011,Ge2023unifying}. 

There has been significant work to close the gap in rheology expression between the two extremes of particle-inertia-dominant and fluid-viscosity-dominant regimes. In addition to time-scale ratios, they are often instead expressed in terms of stress ratios. $I_c$ is proportional to the square root of Bagnold's classic dispersive, or collisional, stress \citep{bagnold1954} [$\tau_c \sim \rho_p (d\dot\gamma)^2$ and $I_c = \sqrt{\rho_p(d\dot\gamma)^2/\sigma}$]. $I_v$ is proportional to a viscous stress ($\tau_v \sim \eta_f \dot\gamma;$ $I_v = \eta_f \dot\gamma / \sigma$). Researchers \citep{pouliquen2002,midi2004,pouliquen2006,jop2006,boyer2011,cassar2005} have demonstrated that a wide range of dry and immersed particle flows can be efficiently expressed in terms of the dependence of $\mu_{\textrm{eff}} \equiv \tau / \sigma$ and the solid fraction $\phi_s$ on the relevant dimensionless number $I_c$ and/or $I_v$ (e.g., Table \ref{table:1}). However, the connection between collision-dominated, viscous-dominated, and (what might be called) visco-collisional rheology \citep{trulsson2012} is still in question. We consider that shear stress in dense particle-fluid flows can be generally expressed as a superposition of $\tau_c$ and $\tau_v$ and that their relative contributions are independent functions of $\phi_s$ [$f_c(\phi_s)$ and $f_v(\phi_s)$]:
\begin{subequations} \label{eq_total_tau}
   \begin{align}
\tau=f_c(\phi_s)\rho_p (d \dot\gamma)^2 + f_v(\phi_s)\eta_f\dot\gamma\ , \label{eq:2a} \\
\mu =f_c(\phi_s)I^2_c +f_v(\phi_s)I_v\ . \label{eq:2b}    \end{align}
\end{subequations}

Initially, we cannot find functional forms for $f_c(\phi)$ and $f_v(\phi)$.  In his classic 1954 paper, \citet{bagnold1954} may have been the first to suggest theoretically and experimentally that we should expect collisional stresses and viscous stresses to increase monotonically with $\phi$, but with different functional dependencies.  Unfortunately, Bagnold did not report any results in which he investigated their effects simultaneously. In contrast, while recent studies have considered the combined effect of collisoinal and viscous stresses [e.g., Refs.\ \citet{cassar2005,trulsson2012}], they suggest $f_c / f_v$ is independent of $\phi_s$. To understand the potential complexity of this problem, we can turn to extensive investigations of meso-scale contact correlations (``fabric'') in quasi-static sheared dry granular flows \citep{bi2011jamming,sarkar2013origin,sarkar2016shear}, in which Berhinger, Chakroborty, and colleagues demonstrated that these evolving structures can arrest flows (referred to as `jamming' the systems) at $\phi_s$'s below which jamming was previously theorized possible.  We suspect that details of this evolving influence the nature of viscous and collisional rate-dependent rheology as well. Recently, \citet{Ge2023unifying} theoretically introduced a new length-scale ratio based dimensionless number, $G$, which was verified using experimental data extracted from \citet{Tapia2022} and \citet{boyer2011}. This new advance links theoretically the viscous-inertial transition of granular-fluid systems to the ratio between microscopic and macroscopic length scales. Thus, it is necessary to investigate the detailed collisional-viscous stress partition in granular-fluid system to clarify the underlying physics behind the regime transition of granular system.

\begin{table}
  \begin{center}
\def~{\hphantom{0}}
  \begin{tabular}{lcc}
      $\ $ & $\mu_{\rm{eff}}$ & $\phi_s$ \\[1ex]
      \hline
       dry beads  & $ \mu_{\rm{1c}}+\Delta\mu_c/(1 + I_{co}/I_c)$ & $\phi_m/(1+\beta_c I_c)$ \\ [1ex] 
       \citep{jop2006} & $\mu_{1c}, \Delta\mu_c, I_{co}=0.265, 0.36, 0.35$ & $\phi_m, \beta_c=0.605,0.25$\\ [1ex]
       \hline
       viscous suspensions  & $\mu_{\rm{1v}}+ \Delta\mu_c/\left(1 + \frac{I_{vo}}{I_v}\right)+I_v+\frac{5}{2}\phi_m \sqrt{I_v}$ & $\phi_m/(1+\beta_v \sqrt{I_v})$\\  [1ex]                      
       \citep{boyer2011} & $\mu_{1v}, \Delta\mu_c, I_{vo}=0.265, 0.36, 0.035$ & $\phi_m, \beta_v=0.605,0.78$\\ [1ex]
       \hline
       dense suspensions & $ \mu_{\rm{1s}}+({\mu_{\rm{2s}}-\mu_{\rm{1s}}})/({1 +\sqrt{I_{so}/I_s}})$ & $\phi_m/(1+\beta_{s1}\sqrt{I_s})$ \\ [1ex]  
        \citep{trulsson2012} & $\mu_{1s}, \mu_{2s}, I_{so}=0.265, 2.2, 0.25$ & $\phi_m, \beta_{s1}=0.605,0.80$\\ [1ex]
        \  & $I_s = I_v +\alpha_{s}I_c$, $\alpha_s\approx 0.03$ & \  \\ [1ex]
        \hline
        dense suspensions & $\mu_c[1+a_{\mu}(I_v+\alpha_{\mu}I_c^2)^{1/2}]$ & $\phi_c[1-a_{\phi}(I_v+\alpha_{\phi}I_c^2)^{1/2}]$ \\ [1ex]
       \citep{Tapia2022} & $\mu_c, \alpha_{\mu}=0.265, 0.0088$ & $\phi_c, \alpha_{\phi}=0.605, 0.1$ \\ [1ex]
        & $a_{\mu}\in [3.79,9.96]$ & $a_{\phi}\in [0.25, 0.42]$ \\ [1ex]
               \hline
       Slurry (lengthscale theory) & $\mu_c(1+\mathcal{A}G^{0.5})$ & $\phi_m(1+b_{\phi}G^{0.5})$ \\ [1ex]  
        \citep{Ge2023unifying} & $\mathcal{A}=a_v-\Delta a/\left(1+cI_v/I_c\right)$ & $\phi_m, b_{\phi}=0.605,0.188$ \\ [1ex]
          & $\mu_{c}, a_v, \Delta a, c = 0.265, 4.0, 3.0, 41$ &  
  \end{tabular}
  \caption{Fits from literature and fit parameters for data in Fig. 2. The equations in this table are from previous literature \citep{jop2006,boyer2011,trulsson2012,Tapia2022,Ge2023unifying} and fitted with our own simulation results. We note that $\mathcal{A}$ in the lengthscale theory proposed by \citet{Ge2023unifying}, instead of being a constant, is a function of $I_c/I_v$. Meanwhile, in \citet{Tapia2022}, either $a_{\mu}$ or $a_{\phi}$ was proposed to be within a range. To fit our own simulation results into this theoretical framework, we use the same range for $a_{\mu}$ as \citet{Tapia2022}, but a different range for $a_{\phi}$.}
  \label{table:1}
  \end{center}
\end{table}

In this paper, we investigate the dynamics of and transitions between collisional, viscous, and visco-collisional flows using 3-D discrete element model (DEM) simulations elaborated in Sec. \ref{sec:method} \citep{cundall1979}. The simulation results indicate that our numerical model, while being simple, is capable of capturing behaviours of granular-fluid systems transitioning from collision-dominant behaviour to viscosity-dominant behaviour. In contrast with previous formulations for collisional flows \citep{jop2006,pouliquen2006} and viscous flows \citep{cassar2005,trulsson2012}, we find shearing these systems under systematic variation of $I_c$ and $I_v$ gives rise to a multi-value relationship between $\mu_{\rm{eff}}$ and $\phi_s$, as suggested by Eqn. \ref{eq:5b}.  Then we obtain relatively simple measures of the fabric to interrogate the nature of the transitions between these different ``phases''. Finally, we use our data to derive expressions for $f_c(\phi_s)$ and $f_v(\phi_s)$ by comparing theoretical forms of collisional and viscous stresses with their numerically obtained counterparts, and further introduce the influence of granular temperatures, before drawing conclusions in Sect. \ref{sec:conclud}.

\section{Methodology}\label{sec:method}

\subsection{Discrete element method}
\label{sec:DEM}
We investigate the behaviour of granular systems from a particle-scale perspective implementing a discrete element method (DEM) simulation with some functional representations of the influence of interstitial fluids. Both the normal and tangential contact forces are calculated based on Hertz-Mindlin contact theories \citep{cundall1979} and damping components based on the derivation outlined by \citet{tsuji1992}. In this model, the tangential contact force also follows the Coulomb friction law, where the tangential contact forces cannot exceed $\mu_p F_{c}^{\rm{ij,n}}$. The magnitude of both normal and tangential contact forces are calculated from
\begin{subequations} \label{eq:contact}
\begin{align}
F_{c}^{\rm{ij,n}} = -k_n\delta_{n}^{1.5} - \eta_n\delta_{n}^{0.25}\dot{\delta}_{n}\ , \label{eq:contacta}\\
F_{c}^{\rm{ij,t}} = \rm{min}\big( -k_{t}\delta_{n}^{0.5}\delta_t - \eta_{t}\delta_n^{0.25}\dot{\delta}_t, \mu_{p}\it{F}_{c}^{\rm{ij,n}} \big)\ , \label{eq:contactb}
\end{align}
\end{subequations}
where $F_{c}^{\rm{ij,n}}$ and $F_{c}^{\rm{ij,t}}$ are normal and tangential contact forces acting on particle \textit{i} from particle \textit{j}. $\delta_{n}$ is the overlap between particles in the normal direction in the DEM simulation, which is given by $\delta_{n} = R_{i}+R_{j} - |\vec{r}_{i} - \vec{r}_{j}|$, where $R_i$ and $R_j$ are particle radii, and $\vec{r}_{i}$ and $\vec{r}_{j}$ are the position vectors of two particles. $\delta_t$ is the corresponding tangential deformation at the contact point between two particles, and $\mu_{p}$ is the coefficient of friction. The coefficients in the contact model are related to material properties of two contacting particles and are presented in Table \ref{table:2}. In this simulation, the particle density is 2650 kg/m$^3$, elastic modulus is 29 GPa, and the Poisson's ratio is 0.20. The material properties are based on those of granite spheres. In order to calculate the dissipative term, the coefficient of restitution is set to be around 0.20. Usually, in studies related to granular materials, the particles are relatively smooth and elastic. However, in this study, we chose a relatively small number for the coefficient of restitution to capture a more representative collisional behaviour of granite sand particles whose collisions are less elastic \citep{man2022two}. According to \citet{foerster1994}, the frictional coefficient between particles, $\mu_p$, is set to 0.1.

\begin{table}
  \begin{center}
\def~{\hphantom{0}}
  \begin{tabular}{lc}
      Variables & Equations \\[3pt]
      \hline
      $k_n$ & $(4/3)\sqrt{R_{\textrm{eff}}}E_{\textrm{eff}}$ \\
      $k_t$ & $8\sqrt{R_{\textrm{eff}}}G_{\textrm{eff}}$ \\
      $\eta_{n}$ & $\alpha_o \sqrt{m_{\textrm{eff}}k_{n}}$ \\
      $\eta_{t}$ & $\alpha_o \sqrt{m_{\textrm{eff}}k_{t}}$ \\
      $R_{\textrm{eff}}$ & $(1/R_i + 1/R_j)^{-1}$ \\
      $E_{\textrm{eff}}$ & $[(1-\nu_{i}^2)/E_i + (1-\nu_{j}^2)/E_j]^{-1}$ \\
      $G_{\textrm{eff}}$ & $[2(1+\nu_{i})(2-\nu_i)/E_i + 2(1+\nu_{j})(2-\nu_j)/E_j]^{-1}$ \\
      $m_{\textrm{eff}}$ & $(1/m_i + 1/m_j)^{-1}$ \\
  \end{tabular}
  \caption{Relationships for calculating the stiffnesses and damping coefficients in Eqn. \ref{eq:contact}. $\alpha_o = 0.9$ is calculated based on the relationship between $\alpha_o$ and $e$ proposed by \citet{tsuji1992}. $E_i$ and $E_j$ are elastic moduli, $\nu_i$ and $\nu_j$ are Poisson's ratios, and $m_i$ and $m_j$ are masses of contacting particles $i$ and $j$.}
  \label{table:2}
  \end{center}
\end{table}

\begin{figure}
  \centering
  \includegraphics[scale = 0.6]{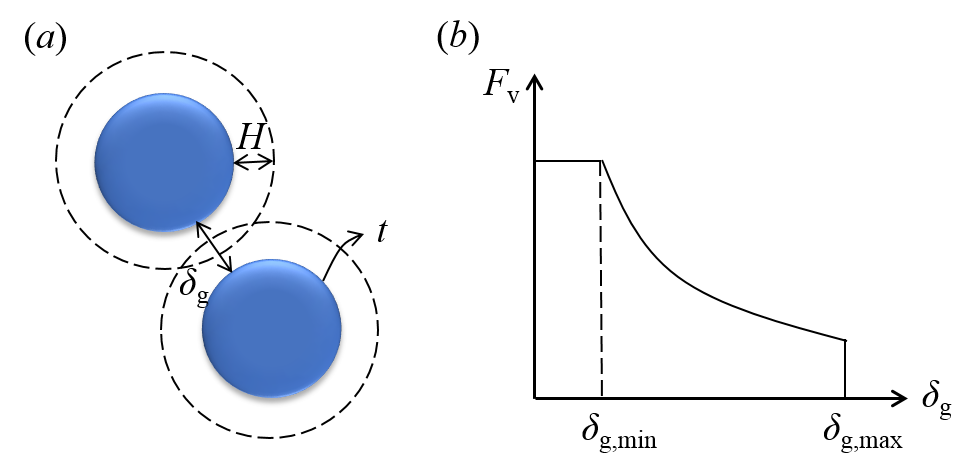}
  \caption{Sketches for considering the lubrication effect between adjacent particles with the surface roughness, $\delta_{g,min}$, and the cut-off length, $\delta_{g,max}$.}
  \label{fig:1}
\end{figure}

To model the influence of properties of a interstitial fluid whose influence is like a thick coating or a spherical lubrication zone surrounding each particle [dashed lines in Fig. \ref{fig:1}(a)], we use a lubrication model suggested by \citet{pitois2000} and \citet{goldman1967} as in Refs.\  \citet{liu2013,marshall2014}. The forces, due to the interstitial fluid between two particles, normal and tangential to their closest surface points can be written using the following equations
\begin{subequations} \label{eq:lub}
\begin{align}
F_{\rm lub}^{\rm{ij,n}} = 6\pi\eta_f R_{\rm{eff}}^{2}G_{f}^{2}\left(v_{n}^{\rm{rel}}/\delta_g 
\right)\ , \label{eq:luba}\\
F_{\rm lub}^{\rm{ij,t}} = 6\pi\eta_f R_{\rm{eff}}v_{t}^{\rm{rel}}\left[\frac{8}{15}\rm{ln}(R_{\rm{eff}}/\delta_{g})+0.9588\right]\ , \label{eq:lubb}
\end{align}
\end{subequations}
where $F_{\rm lub}^{\rm{ij,n}}$ and $F_{\rm lub}^{\rm{ij,t}}$ are normal and tangential lubrication forces between particle $i$ and particle $j$, $R_{\rm{eff}}$ is the effective radius calculated based on the radius of two contacting particles, $v_{n}^{\rm{rel}}$ and $v_{t}^{\rm{rel}}$ are relative normal velocity and relative tangential velocity, respectively, $\eta_f$ is the fluid viscosity, and $\delta_g$ is the gap between the nearest surface of two particles. $G_f$ is a coefficient considering the effective fluid volume which has a lubrication effect to make sure that the influence of lubrication will be decreased as we increase the gap between adjacent particles. 

As shown in Fig. \ref{fig:1}(b), we use two regularizing length scales for this viscous force: (1) $\delta_{\rm{g,min}}\sim \langle d \rangle /10 \approx 2t$ can be thought of as twice a typical particle roughness length scale, $t$; and (2) $\delta_{\rm{g,max}} \sim \langle d \rangle \approx 2H$ (twice the maximum lubrication length scale, $H$). In this study, we consider thick spherical lubrication zones on the surface of particles. Thus, when the distance between two particle is too large, the lubrication effect is negligible. We choose average particle diameter as the maximum lubrication length scale to restrict the lubrication forces to those between two particles only when they are reasonably close. For instance, when the distance of two particles is larger than a particle diameter, another particle may get into the gap between two particles, thus, the lubrication effect should not be considered. Fig. \ref{fig:1}(b) shows the schematic relationship between lubrication forces, $F_{l}^{\rm{ij,n}}$ or $F_{l}^{\rm{ij,t}}$, and particle gap, $\delta_g$, keeping the relative particle velocities constant. $G_{f}$ is calculated using the equation
\begin{equation}
    \begin{split}
        G_{f} = 1 - \left({1+\frac{\bar{V}}{\pi R_{\rm{eff}}\delta_{g}^2}}\right)^{-1/2} \ , \label{Gf}
    \end{split}
\end{equation}
where $\bar{V}$ is the effective lubrication volume, and $\bar{V} = [\delta_{\textrm{g,max}}/(1+0.5\theta_c)]^3$. Here, we consider $\delta_{\rm{g,max}}$ as a rupture distance when no lubrication effects exist between two adjacent particles. $\theta_c$ is the contact angle of the interstitial fluid and associated to the surface energy of the interstitial fluid. In this simulation, we set $\theta_c$ to be constant and equal to 0.5.

\subsection{Simulation setup}
\label{sec:setup}
\begin{figure}
	\centering
	\includegraphics[scale = 0.52]{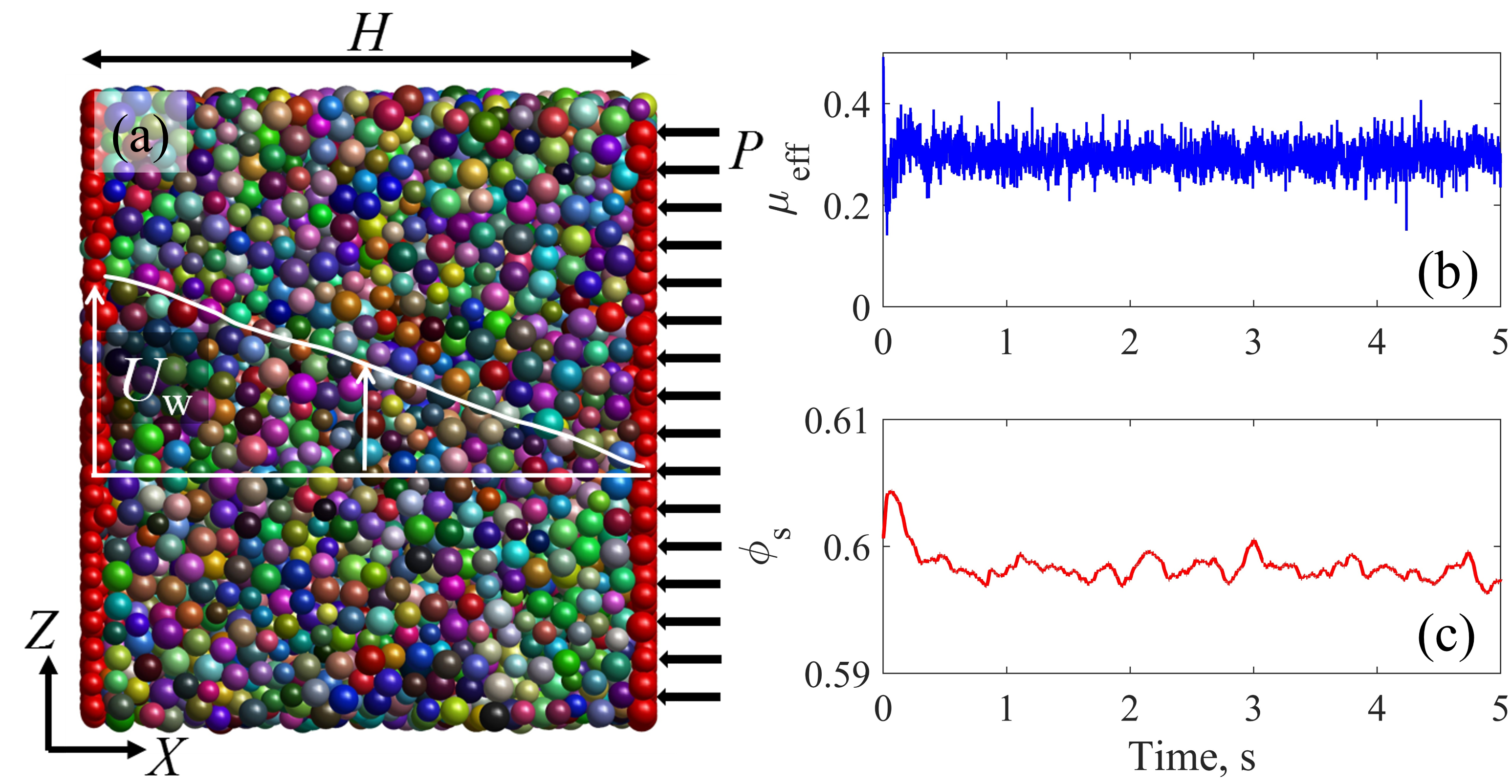}
	\caption{(a) Simulation set-up, which also shows the typical velocity profile when the granular system is being sheared. (b) Relationship between the simulation time and the effective frictional coefficient, $\mu_{\rm eff}$. (c) Relationship between the simulation time and the average solid fraction, $\phi_s$, of the system. We note that the granular system we are testing is three dimensional. We did not show the thickness in $Y-$direction for simplification reasons.}
\label{fig:3}
\end{figure}

To measure the rheology of granular-fluid systems, we shear them in a rectangular box (Fig. \ref{fig:3}(a)) with periodic boundary conditions in the $Y$- and $Z$- directions and roughened walls (using ``glued'' particles) in the $X$-direction. We note that the shearing velocity is along the $Z$-direction shown in Figure \ref{fig:3}. The granular system consists of 6400 particles with a uniform size distribution from 2.0 mm to 3.0 mm and an average particle diameter of 2.5 mm. To shear the granular assembly, we move one of the rough walls in the $Z$-direction (only) with a constant velocity $U_w$. We apply a constant normal stress ($\sigma$) to the other roughened wall and allow it to move (only) in the $X$-direction. We run this simulation until the system reaches a statistically steady state (Fig. \ref{fig:3}(b,c)) and calculate the steady state average values of shear stress $\tau$, shear rate $\dot{\gamma}$, solid fraction $\phi_s$, and effective frictional coefficient $\mu_{\rm eff}$, so that we can also calculate the average values of both the inertial number $I_c = \dot{\gamma}d/\sqrt{\sigma/\rho_p}$ and the viscous number $I_v = \eta_f\dot{\gamma}/\sigma$. 

\begin{table}
  \begin{center}
\def~{\hphantom{0}}
  \begin{tabular}{cccccc}
      $\langle d\rangle \pm \sigma_d $ & $\rho_p $ & $\eta_f$ & $\sigma$ & $ u_h$ & $\dot{\gamma}=U_w / h$  \\ [0.5ex]
      (mm) & (kg/m$^3$) & (cP) & (Pa) & (mm/s) & (1/s)  \\ [0.5ex]
      \hline
      $1.25 \pm 0.25$ & 2650 & $ 0,\ 10^{- 4}-10^{4}$ & $100 - 500$ & $10 - 10^{3}$ & $0.16 - 46 $ \\ [1ex] 
  \end{tabular}
  \caption{Input Parameters. We note that $\sigma_d$ denotes the variation of the particle diameter.}
  \label{table:3}
  \end{center}
\end{table}

\subsection{Model validations}
\label{sec:valid}

We perform ten sets of simulations designed to vary the shearing velocity $U_w$, pressure $\sigma$, and the viscosity of the interstitial fluid $\eta_f$, so that $I_c$ will change from approximately $10^{-3}$ to $1$ and $I_v$, when non-zero, from $10^{-7}$ to $1$. As stated in Sect. \ref{sec:DEM}, we set the elastic modulus of the particle to be $E = 29$ GPa, the coefficient of restitution to be 0.2, and the Poisson ratio to be $\nu = 0.2$. \citet{man2023friction} investigated the influence of inter-particle friction $\mu_p$ on the rheological behaviour of dry granular system and showed that the interparticle friction plays an important role in the macroscopic rheology of the ensemble. However, in this study, we choose to keep $\mu_p$ constant. 

\begin{figure}
	\centering
	\includegraphics[scale = 0.55]{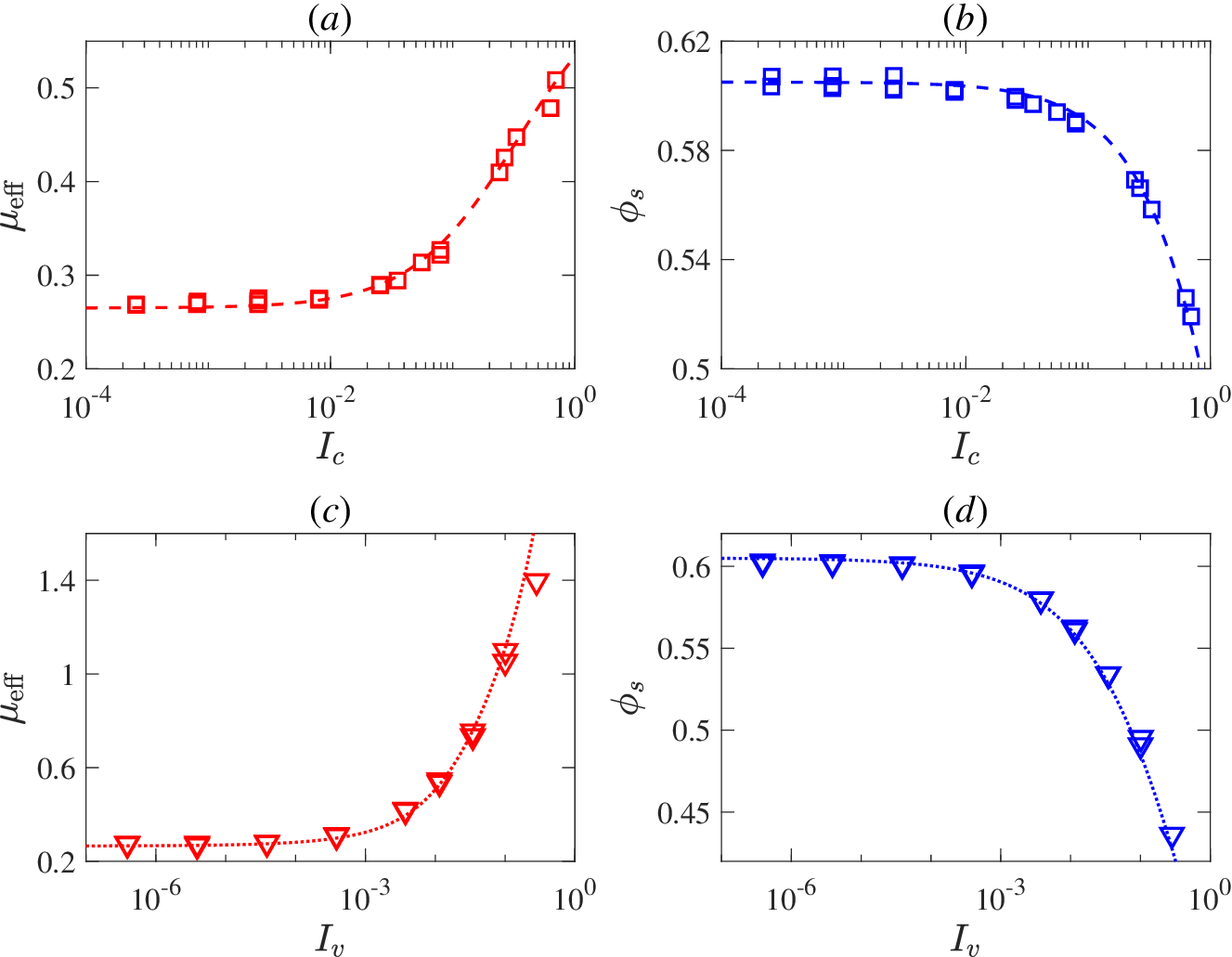}
	\caption{Simulation results for dry granular systems: (a) relationship between $\mu_{\rm{eff}}$ and $I_c$ with a fitting curve of $\mu_{\rm{eff}} = 0.265 + (0.625-0.265)/(1 + 0.35/I_c)$ (red dashed curve) and (b) relationship between $\phi_{\rm{s}}$ and $I_c$ with a fitting curve of $\phi_{\rm{s}} = 0.605/(1+0.25I_c)$. Simulation results for systems with $I_c < 0.01$, where the viscous effect dominates: (c) relationship between $\mu_{\rm{eff}}$ and $I_v$ with a fitting curve of $\mu_{\rm{eff}} = 0.265 +$ $0.36/(1 + 0.035/I_v) + I_v + 2.5\times 0.605\sqrt{I_v}$ (red dotted curve) and (d) relationship between $\phi_{\rm{s}}$ and $I_v$ with a fitting curve of $\phi_{\rm{s}} = 0.605/(1+0.78\sqrt{I_v})$.}
\label{fig:4}
\end{figure}

In particular, we first examine the behaviour of granular systems with $\eta_f = 0$ cP, i.e., dry granular systems, to show some preliminary results. Fig. \ref{fig:4}(a,b) plots both the relationship between $\mu_{\rm{eff}}$ and $I_c$ and the relationship between $\phi_{\rm{s}}$ and $I_c$. Similar to the classical results of dry granular systems, as we increase $I_c$ from 10$^{-4}$ to 1, $\mu_{\rm{eff}}$ increases nonlinearly from $\approx 0.26$ to $\approx 0.5$, while $\phi_{\rm{s}}$ decreases from 0.61 to 0.53. Thus, for a constant pressure simulation, increasing $I_c$ leads the granular system to shift from a dense system to a semi-dilute system. The $\mu_{\rm{eff}}-I_c$ and $\phi_s-I_c$ relationships can be fitted using equations with classical $\mu-I$ and $\phi-I$ functional forms as shown in the second row of Table \ref{table:1} for the dry systems\citep{jop2005}. 

We also tested the rheological behavior when the inertial number, $I_c$, of a granular system is small. In Fig. \ref{fig:4}(c,d), we plotted the results for systems with $I_c < 0.01$ while imposing a wide range of $I_c$ from $10^{-7}$ to $0.3$. We can see that increasing $I_v$ can also cause the granular system to shift from a dense system to a semi-dilute system. Both the $\mu_{\rm{eff}}\sim I_v$ and $\phi_s\sim I_v$ relationships can be fitted by the equation proposed by \citet{boyer2011} shown in Tabel \ref{table:1}. Based on these preliminary simulations, we further change the viscosity of the interstitial fluid to obtain sheared systems with different combinations of the inertial and viscous numbers [$I_c \in (10^{-3}, 1)$ and $I_v \in (10^{-7}, 1)$].

\begin{figure}
	\centering
	\includegraphics[scale = 0.48]{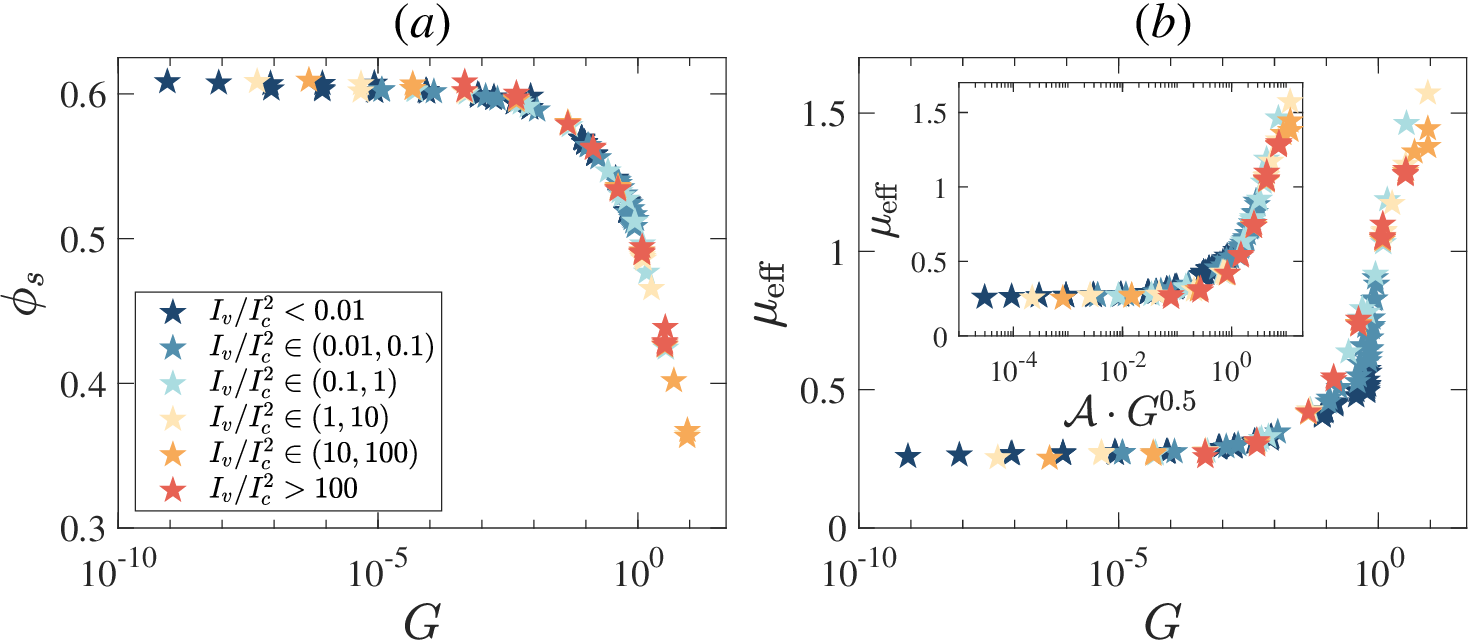}
	\caption{Our simulation results plotted with the dimensionless number, $G$, proposed by \citet{Ge2023unifying}: (a) The relationship between $G$ and $\phi_s$, and (b) the relationship between $G$ and $\mu_{\rm{eff}}$. The inset of Fig. (b) shows that relationship between $\mathcal{A}G$ and $\mu_{\rm{eff}}$, where $\mathcal{A}$ is a function of $I_v/I_c$. We note that \citet{Ge2023unifying} verified the derivation of $G$ based established experimental works conducted in \citet{boyer2011} and \citet{Tapia2022}.}
\label{fig:5}
\end{figure}

Since the theoretical work of \citet{Ge2023unifying} is based on experimental data of \citet{boyer2011} and \citet{Tapia2022}, we incorporate our own DEM-lubrication simulation data within their theoretical framework to investigate the applicability of $G$ in our simulation work and to further validate the credibility of our own numerical model. As a brief review, \citet{Ge2023unifying} proposed a new length-scale based dimensionless number, $G$, which can be specified as
\begin{subequations} \label{eq:GePRF}
    \begin{align}
        G = 12\left(I_v + \lambda_{\rm{St}}I_c^2\right)\ , \label{eq:Gnumber} \\
        \lambda_{\rm{St}} = \frac{1-\exp(-18/\rm{St})}{18-\rm{St}\cdot\left[ 1-\exp(-18/\rm{St}) \right]} \ , \label{eq:lambda}
    \end{align}
\end{subequations}
where $\rm{St}=\it I_c^2/I_v$ is the granular Stokes number. The analytically obtained dimensionless number, $G$, suggests that the combination factor of $I_v$ and $I_c^2$ changes with respect to the ratio between $I_v$ and $I_c^2$, which is slightly different from \citet{Tapia2022} where $\lambda_{\rm{St}}$ is fitted without being linked to neither the intrinsic material properties nor the extrinsic loading conditions, but is quite similar to our argument that the combination of $I_v$ and $I_c^2$ is nonlinear (Sec. \ref{sec:transition}). We first plot $\phi_s$ of our simulation results against $G$ in Fig. \ref{fig:5}(a). This results in a almost perfect collapse of all the simulation data, which is the same as the data collapse in \citet{Ge2023unifying} with the data compile from \citet{boyer2011} and \citet{Tapia2022}. We then plot the relationship between $\mu_{\rm{eff}}$ and $G$ in Fig. \ref{fig:5}(b). Similar to the $\mu_{\rm{eff}}-G$ relationship in \citet{Ge2023unifying}, $\mu_{\rm{eff}}$ cannot be determined by $G$ alone, and we need another variable to collapse this set of data. \citet{Ge2023unifying} suggested that $G$ should be multiplied by a function of $I_c/I_v$ and also provided a functional form of it as follows,
\begin{equation}
    \begin{split}
        \mathcal{A} = f_a\left( I_c/I_v \right) = a_v - \Delta a/\left( 1 + c\cdot I_v/I_c \right), \label{eq:mathcalA} 
    \end{split}
\end{equation}
where $a_v = 4$, $\Delta a = 3$ and $c=41$ are fitting parameters obtained by \citet{Ge2023unifying} based on the experimental results of \citet{boyer2011} and \citet{Tapia2022}. We keep the values of these parameters and plot the relationship between $\mathcal{A}G^{0.5}$ and $\mu_{\rm{eff}}$ in the inset of Fig. \ref{fig:5}(b). It shows that, after rescaling $G$, the $\mu_{\rm{eff}}-\mathcal{A}G^{0.5}$ relationship collapses nicely onto one curve, which implies that, unlike $\phi_s$ being a monotonous function of $G$, $\mu_{\rm{eff}}$ is related to both the length scale ratio, $G$, and the ratio between the inertial number and the viscous number, $I_c/I_v$. The collapse of our simulation data onto monotonic curves for both the $\phi_s-G$ relationship and the $\mu_{\rm{eff}}-\mathcal{A}G$ relationship implies that our numerical model, while simple, is capable of capturing the behaviours of granular-fluid systems transitioning from a collision-dominant behaviour to a viscosity-dominant behaviour.

\section{Results and Discussions}\label{sec:res}

\subsection{Frictional rheology} \label{sec:fric_rheo}

As we have stated in the introduction, for dry granular materials, the rheology is controlled by the inertia number, $I_c$, and for dense granular suspensions or other granular materials in the viscous regime, the behaviour is governed by the viscous number, $I_{v}$. However, for a system subjected to a wide range of shear rates and pressures, it is impossible to use only one dimensionless number to describe the rheological behaviour of granular-fluid systems. Thus, different combinations of $I_c$ and $I_v$ were proposed accordingly \citep{trulsson2012,amarsid2017viscoinertial,Tapia2022,Ge2023unifying}. In this work, however, we first examine the influence of $I_c$ and $I_v$ separately.

\begin{figure}
	\centering
	\includegraphics[scale = 0.42]{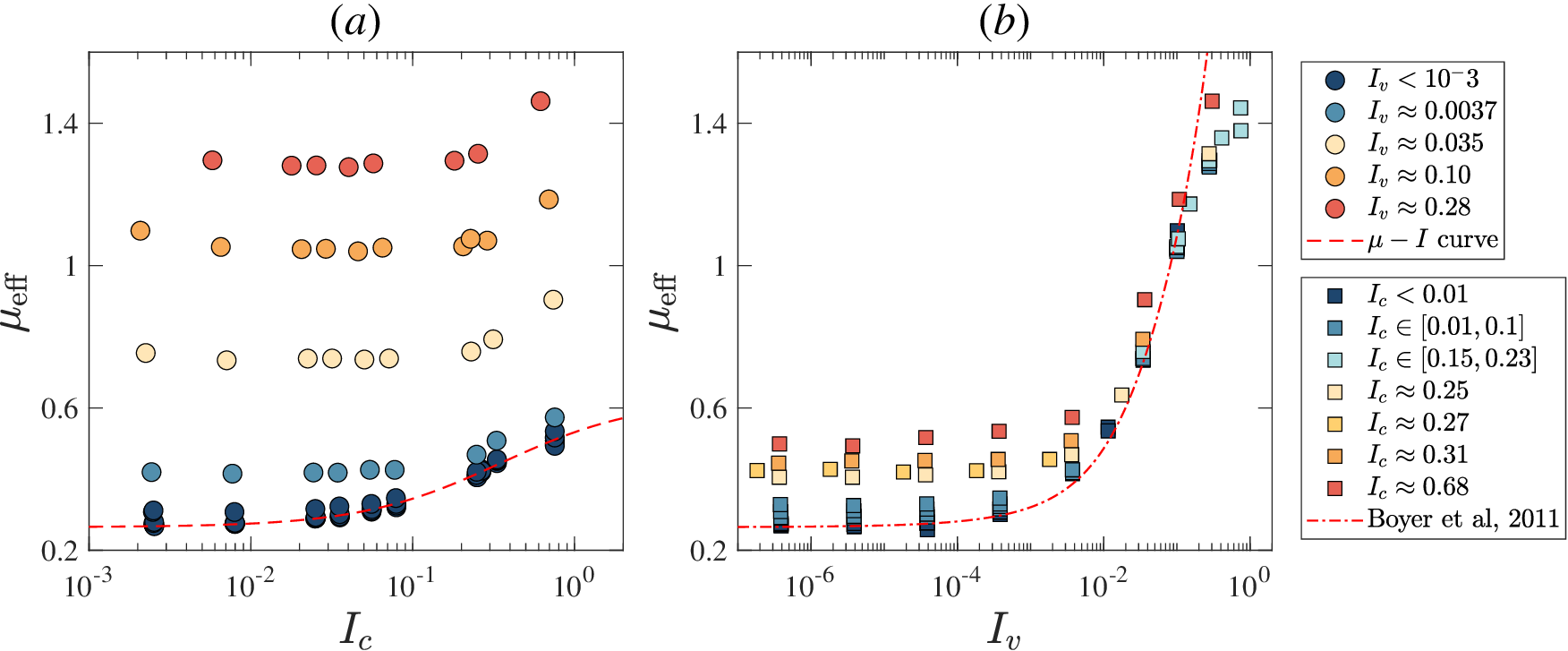}
	\caption{(a) Relationship between $\mu_{\rm{eff}}$ and $I_c$ with different viscous numbers $I_v$. As we increase $I_v$, the $\mu_{\rm{eff}} - I_c$ relationship moves upward. The red dashed curve denotes the classical $\mu-I$ relationship proposed by \citet{midi2004} and with fitting parameters shown in Table \ref{table:1}. (b) Relationship between $\mu_{\rm{eff}}$ and $I_v$ with different viscous numbers $I_c$. As we increase $I_c$, the $\mu_{\rm{eff}} - I_v$ relationship also moves upward, but the amount of increment is not so large as that of the $\mu_{\rm{eff}} - I_c$ curve caused by increasing $I_v$. The red dash-dot curve denotes the $\mu-I_v$ relationship proposed by \citet{boyer2011} and with fitting parameters shown in Table \ref{table:1}.}
\label{fig:6}
\end{figure}

Figure \ref{fig:6}(a) shows the relationship between $I_c$ and $\mu_{\textrm{eff}}$. Different from dry granular materials, the effective frictional coefficients do not have one-on-one relationships with the inertia number any more due to the existence of the interstitial fluid and the lubrication effect introduced by the interstitial fluid. However, the $\mu_{\textrm{eff}} - I_c$ relation for each $I_{v}$ shows similar relationships. This indicates that the effect of particle inertia and the lubrication on the effective frictional coefficient may be separable. Thus, the total friction law of the granular-fluid system may be a linear combination of the contact effect and the lubrication effect. We can see from the figure that, when $I_{v}$ is small ($I_{v} < 10^{-3}$), all the data lie on the same curve [red dashed curve in Fig. \ref{fig:6}(a)]. That is when the particle inertia and particle contact dominant the bulk behaviour of the granular materials. Then, as we increase the viscous number by changing pressure, shear rate or viscosity, the lubrication effect starts to take control. 

Figure \ref{fig:6}(b) shows the relationship between the viscous number, $I_v$, and the effective frictional coefficient, $\mu_{\textrm{eff}}$. Similar to the $\mu_{\textrm{eff}} - I_c$ relationship, as we increase $I_v$, $\mu_{\textrm{eff}}$ is increased accordingly. When, $I_c$ is small, the simulation data almost lie on one curve, which indicates that the inertia number has almost no effect on the behaviour of the granular material. According to the research done by \citet{boyer2011} when they were studying the behaviour of dense granular suspensions in viscous regime, the effective frictional coefficient can be described by a function of the viscous number as shown in the third row in Table \ref{table:1}. The fitting curve is also plotted as the red dash-dot curve in Fig. \ref{fig:6}(b). However, when we increase $I_c$, the simulation results start to deviate from the curve, which indicates that the granular-fluid system starts to be influenced by the inertial effect (or collisional effect).

\begin{figure}
	\centering
	\includegraphics[scale = 0.42]{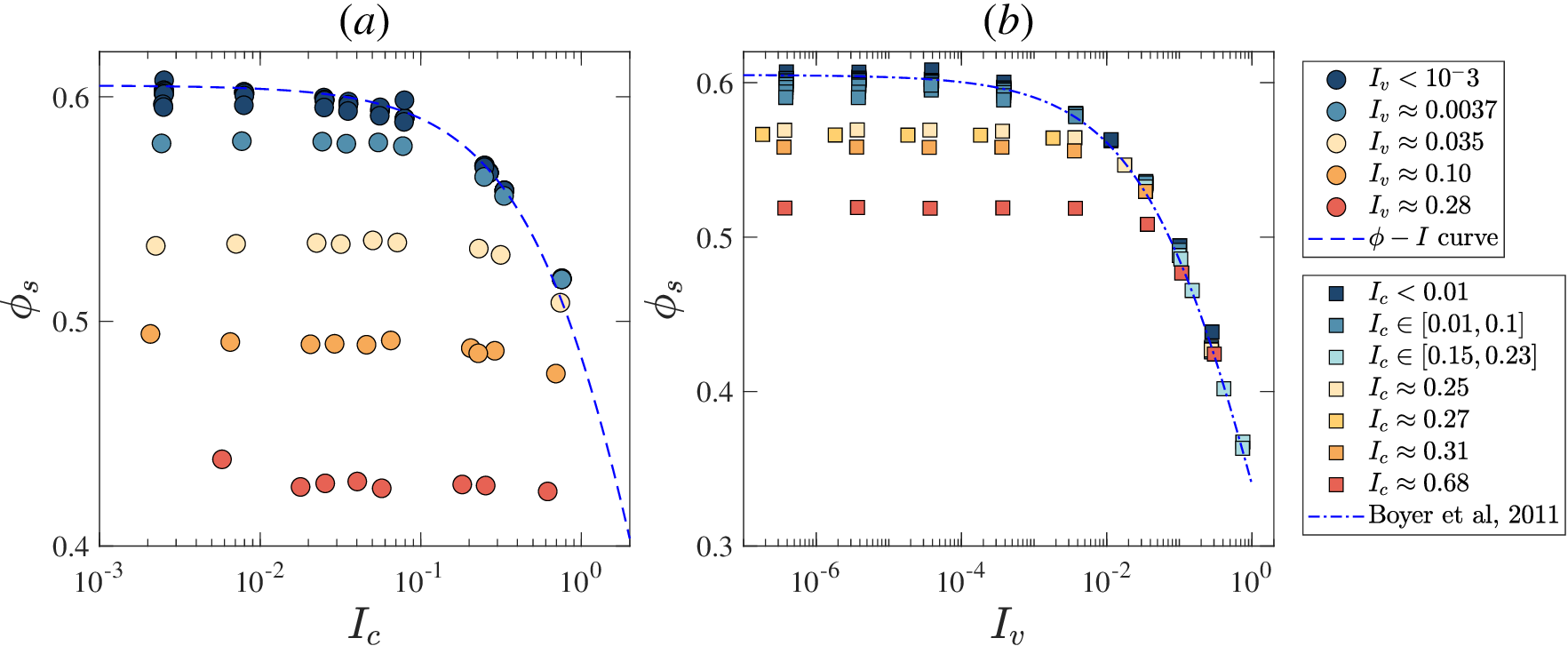}
	\caption{(a) Relationship between $\phi_s$ and $I_c$ with different viscous numbers $I_v$. As we increase $I_v$, the $\phi_s - I_c$ relationship moves downward. The blue dashed curve denotes the classical $\phi-I$ relationship proposed by \citet{midi2004} and with fitting parameters shown in Table \ref{table:1}. (b) Relationship between $\phi_s$ and $I_v$ with different viscous numbers $I_c$. As we increase $I_c$, the $\phi_s - I_v$ relationship also moves upward, but the amount of increment is not so large as that of the $\phi_s - I_c$ curve caused by increasing $I_v$.  Also, when $I_v\gtrapprox 0.1$, the influence of changing $I_c$ is almost negligible. The red dash-dot curve denotes the $\phi-I_v$ relationship proposed by \citet{boyer2011} and with fitting parameters shown in Table \ref{table:1}.}
\label{fig:7}
\end{figure}

For the solid fraction of sheared granular systems, Fig. \ref{fig:7} shows the relationship between the solid fraction of particles and two dimensionless numbers. As we see from Fig. \ref{fig:7}(a), when $I_{v}$ is small, as we increase inertia number, $I_c$, the solid fraction, $\phi_{s}$, is decreased accordingly. We plot the classical $\phi_s-I_c$ relation shown in the second row of Table \ref{table:1} in Fig. \ref{fig:7}(a) as a blue dashed curve. However, when $I_{v}$ is large, the solid fraction is somewhat not influenced by the inertia number, which indicates that, within this regime, the influence of viscosity of the interstitial fluid becomes more important to the behaviour of the granular-fluid system. Figure \ref{fig:7}(b) shows the relationship between $\phi_{s}$ and $I_{v}$. For all cases with different inertia number, the solid fraction decreases with respect to the increase of the viscous number. \citet{boyer2011} proposed that $\phi_{s}$ scales with $(1+I_{v}^{0.5})^{-1}$. We plotted the fitted curve as shown in the third row in Table \ref{table:1} based on the equation proposed by Boyer et al., which is the blue dash-dot curve on Fig. \ref{fig:7}(b). As we can see from the figure, this proposed equation can only predict the cases when $I_c$ is small. 

We note that the comparison between our simulation results and the visco-inertial rheology proposed by \citet{trulsson2012} is also conducted. Given fitting parameters shown in the last row of Table \ref{table:1}, all the simulation results of either the $\mu_{\rm{eff}} - I_s$ relation or the $\phi_s-I_s$ relation collapse onto one curve, which suggests that the visco-inertial rheology of slurries works quite well with our simulation data. However, the feasibility of this rheology highly depends on how we combine $I_c$ and $I_v$. In \citet{trulsson2012}, the combined dimensionless number is $I_s = I_v + \alpha I_c$ with $\alpha \approx 0.635$, but our simulation data yield $\alpha\approx 0.035$. We find that the fitting of $\alpha$ depends on the range of $I_v/I_c^2$, which can be seen as an inverse Stokes number St$^{-1}$, we set up in the simulation, and it is sensitive to the maximum $I_c$ ($I_{c,max}$) we have in the system (which is also confirmed by \citet{Ge2023unifying}). We have addressed that, with this model framework and its corresponding simulation data, it is difficult to combine linearly the inertial number and the viscous number, but we can still find a way to combine the inertial effect and the viscous effect on a macroscopic level. 

\begin{figure}
	\centering
	\includegraphics[scale = 0.42]{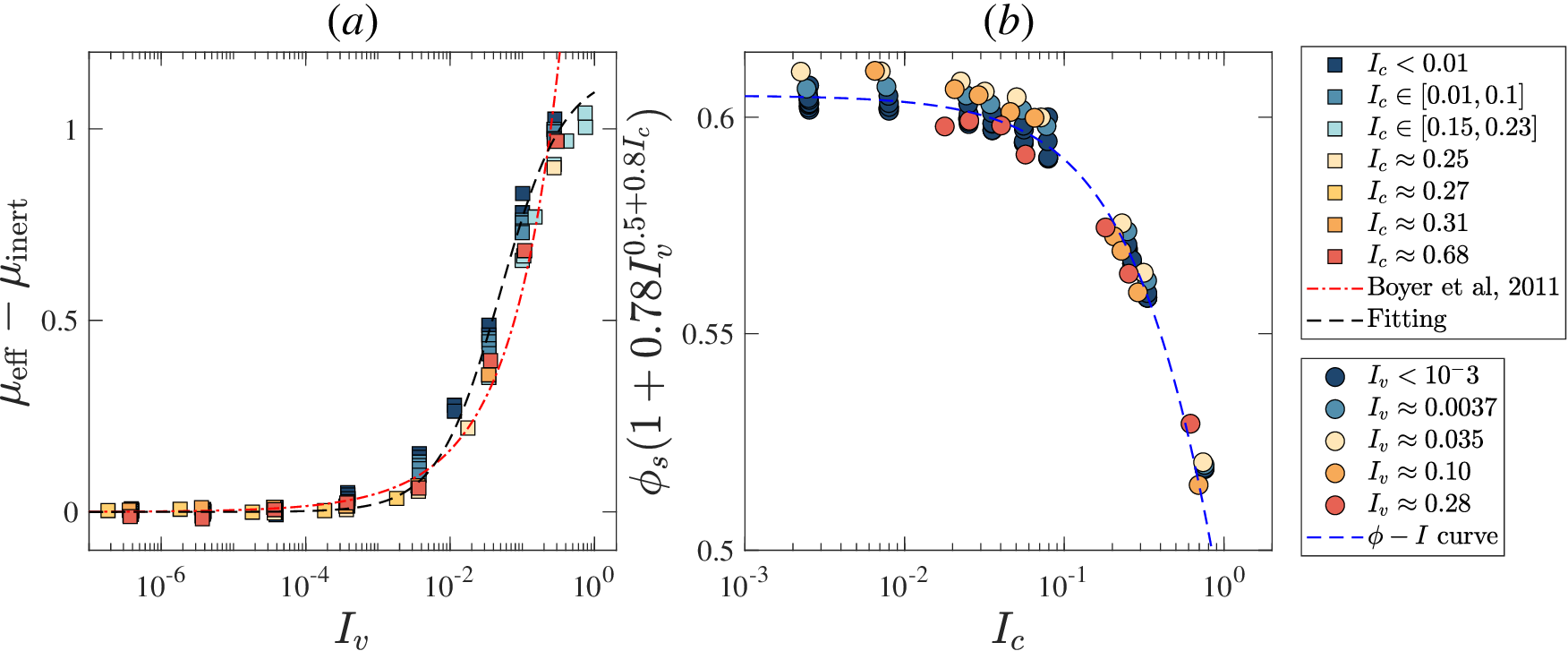}
	\caption{Scalings of $\mu_{\rm{eff}}$ and $\phi_s$: (a) relationship between $\mu_{\rm{eff}}-\mu_{\rm{inert}}$ and $I_v$, and $\mu_{\rm{inert}} = \mu_{\rm{1c}} + (\mu_{\rm{2c}} - \mu_{\rm{1c}})/(1 + I_{\rm{co}}/I_c)$, where $\mu_{\rm{1c}} = 0.265$, $\mu_{\rm{2c}} = 0.625$, and $I_{\rm{co}} = 0.35$. Two fitting curves are $\mu_{\rm{eff}}-\mu_{\rm{inert}} = I_v + 2.5\phi_m I_v^{0.5}$ \citep{boyer2011} and $\mu_{\rm{eff}}-\mu_{\rm{inert}} = \mu_{\rm{3v}}/(1+I_{\rm{v0}}/I_v)$. (b) relationship between $\phi_s(1+\beta_vI_v^{0.5+I_c})$ against $I_c$, where the fitting curve is the same as Eqn. \ref{eq:phis}. When the inertia number is large, the results start to deviate from the prediction.}
\label{fig:8}
\end{figure}

Previous research \citep{boyer2011} suggests that, for the $\mu_{\rm{eff}} = \mu_{\rm{eff}}(I_c, I_v)$ relationship, $\mu_{\rm{eff}}$ can be written as a function of $I_v$ alone, shown as the red dash-dot curve in Fig. \ref{fig:6}(b). However, based on Fig. \ref{fig:6} and its corresponding discussion above, we believe that the influence of particle collision and interstitial fluid, i.e., the influence of $I_c$ and $I_v$, can be separated. Thus, we divide $\mu_{\rm{eff}}$ into two parts: (i) the inertial part that follows the classical dry granular $\mu-I$ rheology that $\mu_{\rm{inert}} = \mu_{\rm{1c}} + (\mu_{\rm{2c}} - \mu_{\rm{1c}})/(1 + I_{\rm{co}}/I_c)$, where $\mu_{\rm{1c}} = 0.265$, $\mu_{\rm{2c}} = 0.625$, and $I_{\rm{co}} = 0.35$, which is the same as that in the second row of Table \ref{table:1}; and (ii) the fluid-influenced part $\mu_{\rm{F}}$ that can be fitted with different forms. We subtract $\mu_{\rm{inert}}$ from $\mu_{\rm{eff}}$ and plot $\mu_{\rm{F}}\equiv\mu_{\rm{eff}} - \mu_{\rm{inert}}$ against $I_v$ in Fig. \ref{fig:8}(a). It shows a good collapse of all the simulation data that $\mu_{\rm{F}}$ can be written as a single function of $I_v$. However, the fitting of the $\mu_{\rm{F}}-I_v$ relationship is debatable, since either the classical functional form of the $\mu-I$ rheology ($\mu_{\rm{F}} = \mu_{\rm{3v}}/(1+I_{\rm{v0}}/I_v)$ and the black dashed curve in Fig. \ref{fig:8}) and the equation proposed by \citet{boyer2011} ($\mu_{\rm{F}} = I_v + 2.5\phi_m I_v^{0.5}$, $\phi_m=0.605$ and the red dash-dot curve in Fig. \ref{fig:8}) can well predict the rheological behaviour. Thus, the $\mu_{\rm{eff}} = \mu_{\rm{eff}}(I_c, I_v)$ relationship can be written as
    \begin{subequations} \label{eq:mueff}
        \begin{align}
            \mu_{\rm{eff}} = \mu_{\rm{1c}} + (\mu_{\rm{2c}} - \mu_{\rm{1c}})/(1 + I_{\rm{co}}/I_c) + \mu_{\rm{3v}}/(1+I_{\rm{v0}}/I_v),\label{eq:mueff1}\\
            \mu_{\rm{eff}} = \mu_{\rm{1c}} + (\mu_{\rm{2c}} - \mu_{\rm{1c}})/(1 + I_{\rm{co}}/I_c) + I_v + 2.5\phi_m I_v^{0.5},\label{eq:mueff2}
        \end{align}
    \end{subequations}
where $\mu_{\rm{1c}}$, $\mu_{\rm{2c}}$, $\mu_{\rm{3v}}=1.1$, $I_{\rm{co}}$, and $I_{\rm{vo}} = 0.05$ are fitting parameters. The advantage of Eqn. \ref{eq:mueff2} is that its fluid-influenced part has no fitting parameters ($\phi_m$ denotes the maximum solid fraction for a shear granular system, which can be seen as a naturally obtained constant) and this functional form naturally converge to the Einstein viscosity for dilute suspensions that $\eta_{\rm{s,d}} = 1 + 2.5\phi_s$. However, for a drained granular-fluid system, its effective frictional coefficient cannot always increase as suggested by Eqn. \ref{eq:mueff2}. When $I_v$ is sufficiently large, $\mu_{\rm{F}}$ may also reach a plateau as suggested by Eqn. \ref{eq:mueff1}. In this aspect, detailed analysis is still needed and we will further utilize a more explicit model framework, such as coupled lattice Boltzmann discrete element method (LBM-DEM), to tackle this challenge.

Similarly, we can also decompose the $\phi_s = \phi_s(I_c, I_v)$ relationship into the influence of particle collisions and that of the interstitial fluid. However, Fig. \ref{fig:7} suggests that this decomposition cannot be done completely. In Fig. \ref{fig:7}(a), as we keep $I_v$ constant and small ($<10^{-3}$) and increase $I_c$, $\phi_s$ will start to decrease from the plateau at $I_c\approx 0.1$. However, when $I_v\approx 0.1$, as we increase $I_c$, $\phi_s$ will not decrease until $I_c\gtrapprox 0.3$, which indicate that, if we regard the solid fraction drop as a regime transition, the transition occurrence is not only governed by $I_c$, but also governed by $I_v$. Fig. \ref{fig:7}(b) also indicates a similar conclusion, but also shows that, interestingly, when $I_v$ is larger than 0.1, the $\phi_s(I_c, I_v)$ relationship can be approximately reduced to the $\phi_s(I_v)$ relationship. To separate the influence of $I_c$ and $I_v$, as we keep $I_c$ constant, we recognize each plateau of the $\phi_s(I_v)$ curve being controlled by $I_c$ alone, which implies that each plateau can be calculated as $\phi_m/(1+\beta_cI_c)$. However, the turning point on the $\phi_s(I_v)$ curve is governed by both $I_v$ and $I_c$, which leads us to the combined equation to describe the $\phi_s = \phi_s(I_c, I_v)$ relationship,
\begin{equation} \label{eq:phis}
    \begin{split}
        \phi_s = \frac{\phi_m}{(1+\beta_cI_c)\left(1+\beta_v I_v^{0.5+I_c}\right)}\ \ ,
    \end{split}
\end{equation}
where $\beta_c = 0.25$ and $\beta_v =  0.78$ are fitting parameters, but the same as those fitted using our simulation data for the $\phi_s(I_c)$ relation of dry granular systems and the $\phi_s(I_v)$ relation of highly viscous suspensions, respectively. We plot $\phi_s(1+\beta_vI_v^{0.5+I_c})$ against $I_c$ in Fig. \ref{fig:8}(b). It shows that all the simulation data collapse onto the dry granular $\phi_s-I_c$ curve with slight deviation when $I_c$ is small. 


\subsection{From collision dominant to viscous dominant}
\label{sec:transition}
\begin{figure}
	\centering
	\includegraphics[scale = 0.38]{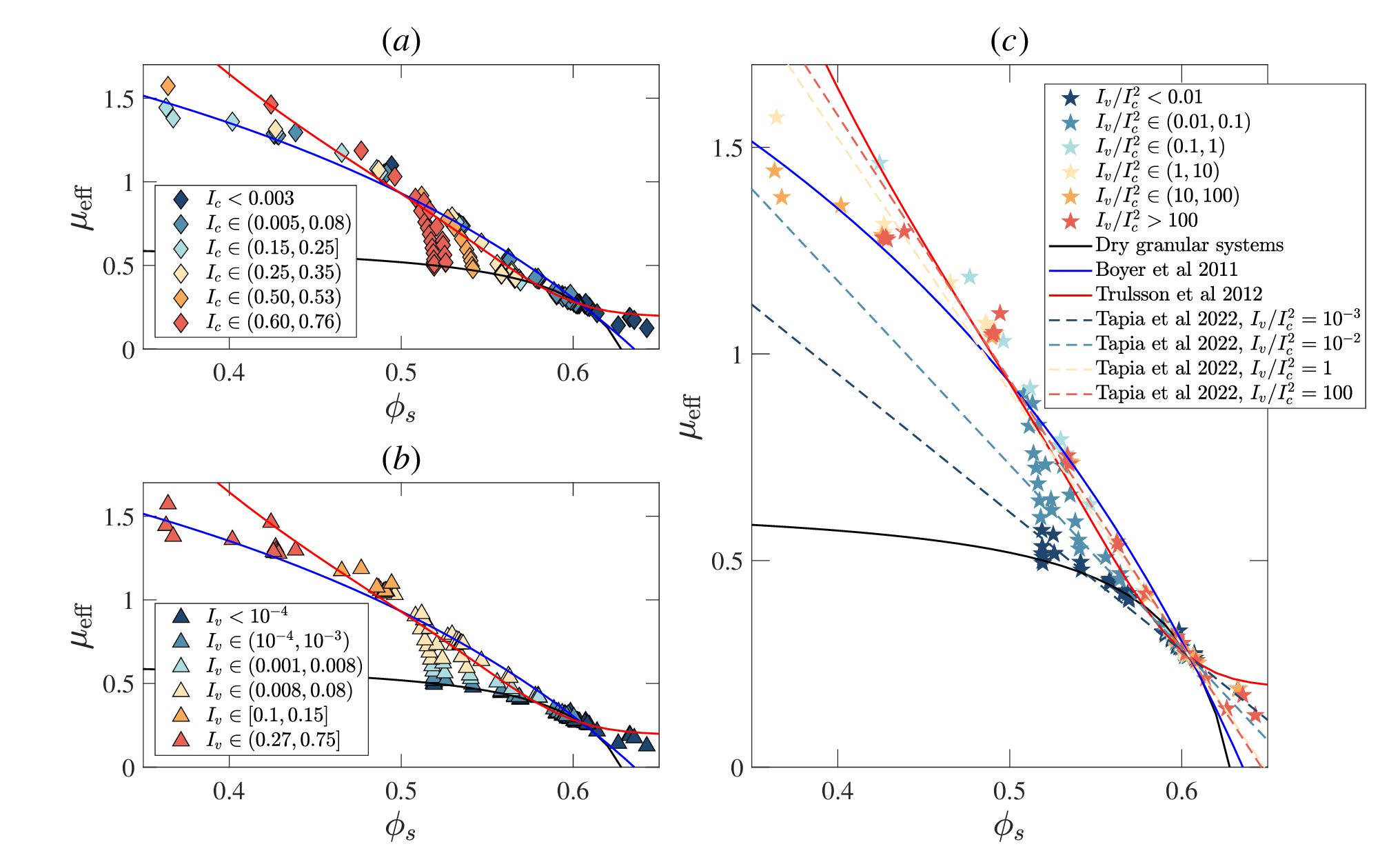}
	\caption{Relationship between $\mu_{\rm{eff}}$ and $\phi_s$ with different control parameters: (a) different colors of markers represent different $I_c$, (b) different colors of markers represent different $I_v$,  (c) different colors of markers represent different St$^{-1} = I_v/I_c^2$.}
\label{fig:9}
\end{figure}

The simulation results and their corresponding discussions suggest that the combination of $I_c-$dominant rheology and $I_v-$dominant rheology is nonlinear. We expect that it is not feasible to write $\mu_{\rm{eff}}$ as a single function of $\phi_s$ as suggested by previous works \citep{pouliquen2006,boyer2011,trulsson2012}. In Fig. \ref{fig:9}, we plot $\mu_{\rm{eff}}$ against $\phi_s$ for all simulation data with different control parameters. Specifically, we examine how the $\mu_{\rm{eff}}-\phi_s$ relationship changes with different ranges of either inertial number $I_c$, viscous number $I_v$, or inversed Stokes number St$^{-1} = I_v/I_c^2$.

Rather than the one-on-one monotonic relationship between $\mu_{\rm{eff}}$ and $\phi_s$, $\mu_{\rm{eff}}(\phi_s)$ is non-monotonic for these system in the range $0.51 \lessapprox \phi_s \lessapprox (0.605 \approx \phi_m)$. $\phi_s \approx 0.51$ corresponds to the lower limit of $\phi$ for the highest value of $I_c$ at which our systems remain uniform. Within this range, the lower bound of $\mu_{\rm{eff}}$ is similar to that previously obtained for collisionally-dominated (dry) flows $\mu_{\rm{eff}}[\phi_s(I_c)]$ \citep{pouliquen2006} (the black solid curve in Fig. \ref{fig:9}). The upper bound for $\mu_{\rm{eff}}$ is dependent on both collisional and viscous effects $\mu_{\rm{eff}}[\phi_s(I_s)]$ (the blue solid curve in Fig. \ref{fig:9}). The upper bound for $\mu_{\rm{eff}}-\phi_s$ can also be fitted by the viscous rheology $\mu_{\rm{eff}}[\phi_s(I_v)]$ proposed by \citet{boyer2011}, shown as the red dashed curve in Fig. \ref{fig:9}. However, the $\mu_{\rm{eff}}[\phi_s(I_v)]$ relationship also implies that, when $\phi_s$ is small, $\mu_{\rm{eff}}$ is approaching infinity. In this range, we transition systems at constant $I_c$ from the lower limit curve $\mu_{\rm{eff}}(\phi_s(I_c))$ to the higher one ($\mu_{\rm{eff}}[\phi_s(I_s)]$ by increasing $\eta_f$ thus increasing $I_v$ (e.g., red arrow in Fig.\ (2(a) for $I_{c,max}$). With an initial increase in $\eta_f$, $\mu_{\rm{eff}}$ increases at a nearly constant $\phi_s$ to the $\mu_{\rm{eff}}[\phi_s(I_s)]$ curve. If we increase $\eta_f$ further still, $\mu_{\rm{eff}}$ increases further while $\phi_s$ decreases along the $\mu_{\rm{eff}}[\phi_s(I_s)]$ curve. 

In more details, Fig. \ref{fig:9}(a) shows that, when we keep $I_c$ almost constant (or within a small interval), increasing $I_v$ helps the system transition from a collision- or inertia-dominant state to a viscous-dominant state. The transition is more evident when $I_c$ is kept at a larger value ($I_c>0.3$). When $I_c < 0.25$, changing $I_v$ will not deviate the $\mu_{\rm{eff}}-\phi_s$ relation away from the visco-inertial rheology proposed by \citet{trulsson2012}. Figure \ref{fig:9}(b), however, provides us with a different perspective on how the $\mu_{\rm{eff}}-\phi_s$ relationship evolves as we stay within different intervals of $I_v$. When $I_v < 0.0008$, the $\mu_{\rm{eff}}-\phi_s(I_c)$ relation follows the classical $\mu-I$ dry granular rheology, but when $I_v>0.1$ and within the viscous regime, the $\mu_{\rm{eff}}-\phi_s(I_c)$ relation is closer to either the viscous rheology proposed by \citet{boyer2011} or the visco-inertial rheology proposed by \citet{trulsson2012}, but their maximum solid fraction, when $I_v$ is kept larger than 0.1, is also much smaller than $\phi_m\approx 0.605$ of the system. Additionally, in Fig. \ref{fig:9}(c), we also examine the behaviour of the $\mu_{\rm{eff}}-\phi_s$ relation as we choose different intervals of the inverse Stokes number, St$^{-1} = I_v/I_c^2$. Following Bagnold's analysis \citep{bagnold1954}, the shear stress of granular systems scales with $\rho_p\dot{\gamma}^2d^2$ in the inertial regime and with $\eta_f\dot{\gamma}$ in the viscous regime. Then, the inverse Stokes number St$^{-1}=I_v/I_c^2=\eta_f\dot{\gamma}/(\rho_p\dot{\gamma}^2d^2)$ can be regarded as the ratio between viscous stresses and inertial-collisional stresses. The overall behaviour of changing St$^{-1}$ is similar to that of changing $I_v$, but it shows a clearer transition from the inertial regime to the viscous regime.

\begin{figure}
	\centering
	\includegraphics[scale = 0.5]{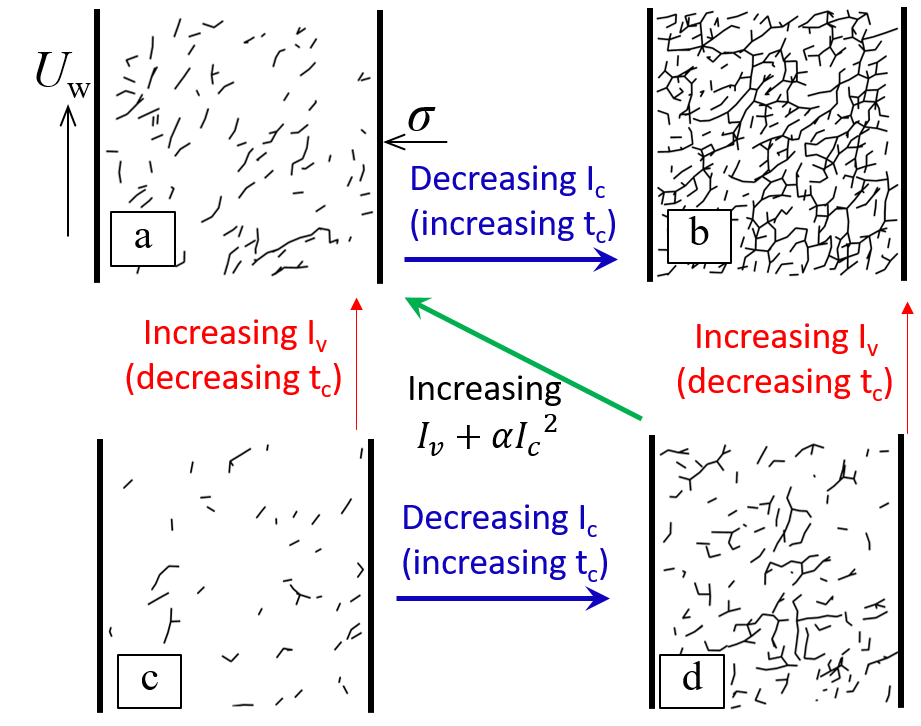}
	\caption{Strong force network and its transition with respect to the change of $I_c$ and $I_v$. Each line connects centers of contacting particles for which the interparticle force $f_{ij}\geq \langle f_{ij} \rangle$, where $\langle f_{ij} \rangle$ is the mean contact force. The combinations of [$I_c$, $I_v$, $\phi_s$] are (upper-left) [0.68, 0.033, 0.51] (lower-left) [0.69, 0.0034, 0.51] (upper-right) [0.0075, 0.012, 0.57] and (lower-right) [0.25, 0.0037, 0.57].}
\label{fig:10}
\end{figure}

To investigate how the fabric evolves under these transitions, for two pairs of experiments with similar $\phi_s$'s and substantially different $\mu_{\rm{eff}}$'s, we plot a representative ``strong force network'' -- line segments for particle pairs for which the interparticle contact force $f_{\rm{cij}}$ is greater than the average (Fig.\ \ref{fig:10}).  For each pair on the same limit curve $\mu_{\textrm{eff}}[\phi(I_c)]$ or $\mu_{\textrm{eff}}[\phi_s(I_s)]$, increasing $\mu_{\rm{eff}}$ simultaneously decreases $\phi_s$ and both connectivity and density of high$-f_{\rm{cij}}$ force pairs. At the same time, enduring extended frictional contact networks are replaced by more isolated collisional contacts. In contrast, for each pair with similar $\phi_s$'s, increasing $\mu_{\rm{eff}}$ limit curve \textit{increases} both connectivity and density of high$-f_{\rm{cij}}$ force pairs. From this, we hypothesize that, as a system transitions from the low limit $\mu_{\rm{eff}}$ curve to the high limit curve at constant $I_c$, modest increases in viscous damping of interparticle movement decreases separation events of contacting particles. Similar to an effective ``stickiness'', this facilitates interparticle connectivity and simultaneously increases $\tau$ relative to $\sigma$ without much change in $\phi_s$. Once the upper $\mu_{\rm{eff}}$ limit curve is reached, further increases in $I_v$ reduce particle contact events, transitioning the system back to one where interparticle contacts are more isolated.

\begin{figure}
	\centering
	\includegraphics[scale = 0.4]{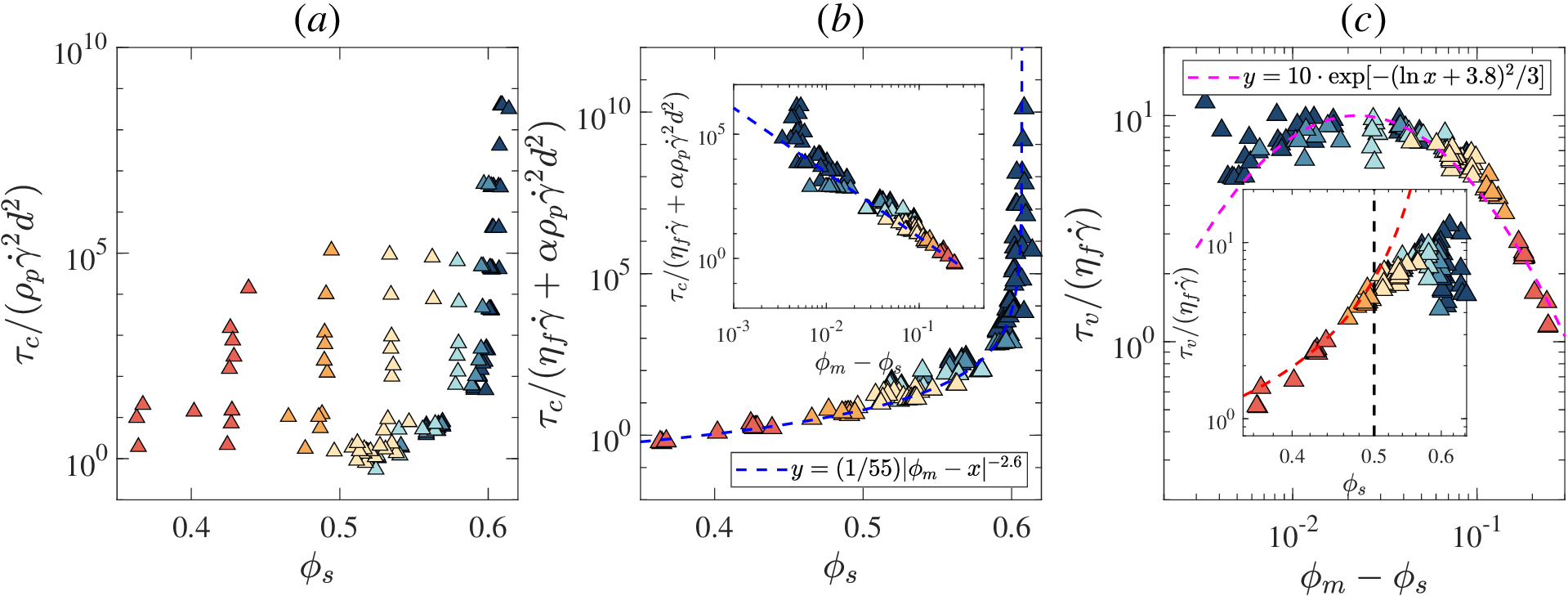}
	\caption{(a) Relationship between solid fraction and the normalized computational contact stress Ratio of computational contract stress $\tau_c/[\alpha \rho_p (d \dot{\gamma})^2]$. (b) Relationship between solid fraction and the normalized computational contact stress Ratio of computational contract stress $\tau_c/[\alpha \rho_p (d \dot{\gamma})^2+ \eta_f \dot{\gamma}]$. The blue dashed line shows that the dimensionless stress scales with $(1/55)(\phi_m - \phi)^{-2.6}$. In the inset, we change the x-axis to $\phi_m - \phi_s$. (c) Ratio of computational theoretical lubrication stresses $\tau_v/(\eta_f \dot{\gamma})$ vs. $\phi_m - \phi_s$. In the inset, we plot The Red dashed line shows that $\tau_v/(\eta_f \dot{\gamma})$ against $\phi_s$, and the red dashed curve denotes that $\tau_v/(\eta_f \dot{\gamma})$ scales with $(1/15)(\phi_m - \phi)^{-2}$ when $\phi_s<0.5$. The markers in this plot are the same as those in Fig. \ref{fig:9}(b).}
\label{fig:11}
\end{figure}

For a quantitative expression of $\mu_{\rm{eff}}$, we calculate contact and viscous shear stresses ($\tau_c$ and $\tau_v$) from the $x$-component of all contact forces and viscous forces between each of our wall particles and adjacent free particles. We find that $\tau_{v} / (\eta_f \dot{\gamma})$, i.e., $\tau_v/(I_v\sigma)$ can be expressed as a single-valued function of $\phi_s$: $f_v(\phi_s)$ [Fig.\ \ref{fig:11}(b)], though $\tau_{c}/[\rho_p (d \dot \gamma)^2]$ cannot [Fig.\ \ref{fig:11}(a), inset]. Rather, $\tau_{c} / [\alpha \rho_p (d \dot \gamma)^2 + \eta_f\dot{\gamma})]$, i.e., $\tau_c/(I_s\sigma)$, is much closer to a single-valued function of $\phi_s$: $f_{c}(\phi_s$) (Fig.\ \ref{fig:11}(a)).  We note that these results suggest that, while non-contact viscous interactions are influenced by $I_v$ alone, contact interactions are influenced by both $I_c$ and $I_v$. Considering the significantly different forms of $f_v(\phi_s)$ and $f_c(\phi_s)$, in place of Eqn.\ (2-3) for granular-fluid systems, we write
\begin{subequations} \label{eq:stresspartition}
\begin{align}
\tau=f_{\textrm{c}}(\phi_s)[\alpha \rho_p (d \dot\gamma)^2 +\eta_f\dot\gamma] + f_{\textrm{v}}(\phi_s)\eta_f\dot\gamma\ , \label{eq:5a} \\
\mu_{\rm{eff}}=f_{\textrm{c}}(\phi_s)\alpha I_c^2 + [f_{\textrm{c}}(\phi_s)+ f_{\textrm{v}}(\phi_s)]I_v\ . \label{eq:5b}
\end{align}
\end{subequations}

Equation \ref{eq:stresspartition} provides an effective, if empirical, expression for the data in Fig.\ \ref{fig:9}.  Further, with Figs.\ \ref{fig:11}, Eqn.\ \ref{eq:5b} provides a strong foundation for intuition into the form of $\mu_{\rm{eff}}(\phi_s)$. For $\phi_s \gtrapprox 0.56$, $f_c / f_v \gg 1$, so for these data $\mu_{\rm{eff}} \approx  f_{\textrm{c}}(\phi_s)I_s$ signifying a regime where viscous and contact effects influence the dynamics.  For $\phi \lessapprox 0.56$ all data on the $\mu_{\rm{eff}}[\phi(I_s)]$ (upper bound) curve in Fig.\ 2(a) correspond to cases where $1 \ll  \alpha I_c^2 / I_v \equiv \rm{St}$ (a Stokes number). For these data $\mu_{\rm{eff}}\approx [f_c(\phi)+ f_v(\phi_s)]I_v$ signifying a regime where viscous effects dominate systems on the upper $\mu_{\rm{eff}}$ curve for $\phi_s \lessapprox 0.56$.  While the functional forms for this upper bound $\mu_{\rm{eff}}$ curve differ, at $\phi_s \approx 0.56$, $I_v \approx I_s$ and $f_{\textrm{c}}(\phi_s)+ f_{\textrm{v}}(\phi_s) \approx f_{\textrm{c}}(\phi_s)$, so there is a smooth transition. Now considering the lower bound $\mu_{\rm{eff}}=\mu_{\rm{eff}}[\phi_s(I_c)]$ curve: for $\phi \lessapprox 0.56$, $\rm{St} \gg 1$ and $f_{\textrm{c}}(\phi_s) \approx f_{\textrm{v}}(\phi_s)$ so $\mu_{\rm{eff}} \approx f_{\rm{c}}(\phi_s)\alpha I_c^2$, dependent primarily on contact effects, as Fig.\ 2(a) and caption imply. For larger $\phi_s$'s on this curve, $I_c$ and St decrease, and the functional form $\mu_{\rm{eff}}=f_{\textrm{c}}(\phi_s)I_s$, the same form as the upper bound curve in this region of $\phi_s$. We propose fitted equations for both $f_c(\phi_s)$ and $f_v(\phi_s)$, but their exact functional forms are less important than the discussions on the granular rheology, thus we only keep them within the caption of Fig. \ref{fig:11}.

\begin{figure}
	\centering
	\includegraphics[scale = 0.45]{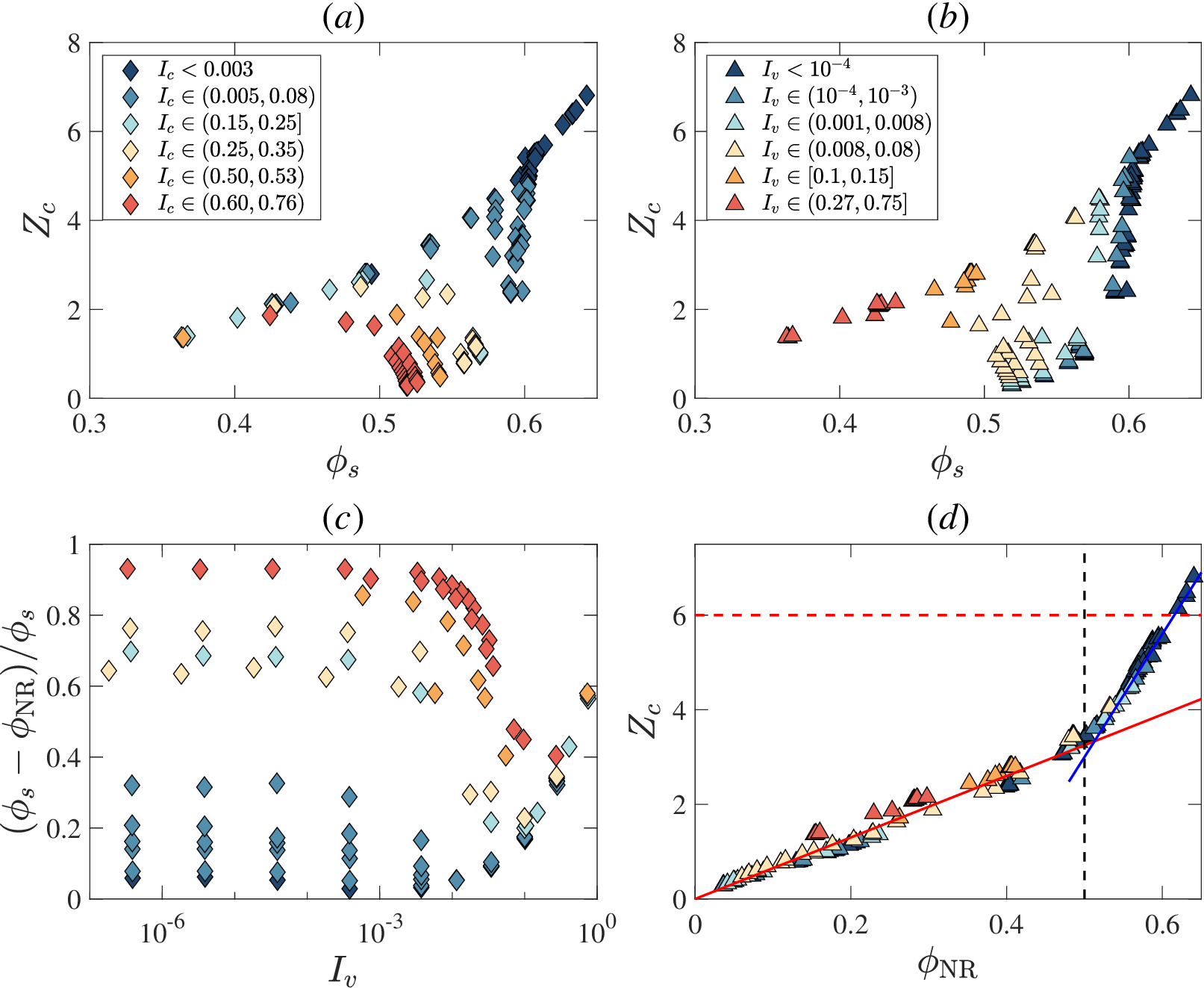}
	\caption{Relationships among coordination number, solid fraction, and non-rattler solid fraction: (a) and (b) plot the relationship between the coordination number $Z_c$ and the solid fraction $\phi_s$. (c) shows the relationship between $(\phi_s-\phi_{\rm{NR}})/\phi_s$ and $I_v$. The markers in Figure (c) are the same as those in Figure (a). (d) Relationship between $Z_c$ and $\phi_{\rm{NR}}$. Markers in Figure (d) are the same as those in Figure (b).}
\label{fig:12}
\end{figure}

This general picture is supported by consideration of the relationship between \textit{coordination number} $Z_c$ (average number of contacts per particle), $\phi_s$, and other bulk parameters (Fig. \ref{fig:12}). As indicated in Figs.\ \ref{fig:12}(a) and (b), for the same moderate increase in $I_v$ for which the system remained primarily on the low $\mu_{\rm{eff}}(\phi_s(I_c))$ limit curve, $Z_c$ maintains a near-constant small value. We also check the fraction of particles potentially available for a connected strong force network [Fig.\ \ref{fig:12}(c)]. $\phi_{\rm{NR}}$ is the part of the system solid fraction composed of particles contacting more than one particle (NR stands for non-rattler particles, as in \citet{bi2011jamming}). In this study, we calculate the non-rattler solid fractions, $\phi_{\rm{NR}}$, by summing the volumes of particles that have more than one contact. In other words, a non-rattler particle has contact numbers $N_c\geq 2$.

We plot the relationship between $\phi_s - \phi_{\rm{NR}}$ and $I_v$ in Fig. \ref{fig:12}(c). For highest $I_c$ and $I_v<0.2$, $F \equiv \phi_s - \phi_{\rm{NR}} \approx 0.93$ indicating that on average only $\approx 7\%$ of the particles have more than one contact, supporting the picture of a system dominated by isolated collisions.  Then $F$ first decreases as $Z_c$ increases (from $I_v \approx$ 0.2 to 0.1), signifying the more highly connected force network.  With additional increase in $I_v$, $F$ rises as $Z_c$ drops again, signifying the return to a less well-connected contact network. When we plot $Z_c$ vs.\ $\phi_{\rm{NR}}$ in Figure \ref{fig:12}(d), the data collapse convincingly. In other words, $Z_c$ and $\phi_{\rm{NR}}$ are much more correlated with system behaviour than is $\phi_s$, as others have found for somewhat different particulate systems \citep{bi2011jamming}. In their 2-d experimental jamming studies, \citet{bi2011jamming} found a similar functional form and slope change in $Z_c$ vs.\ $\phi_{\rm{NR}}$ as we do in our flowing system [Fig.\ \ref{fig:12}(d)] as well as a slope change with a change of fabric structures. Links like this gives insight to a more broadly unified physical framework for dense particle-fluid deformation and flows. When $\phi_{\rm{NR}} \lessapprox 0.5$ ($Z_c \lessapprox 3.0$), the collapsed data forms a linear relationship where $Z_c = 6.5\phi_{\rm{NR}}$. When $\phi_{\rm{NR}}>0.5$ and $Z_c > 3.0$, the slope of the data starts to increase and the final slope becomes 26, which is exactly four times of the scaling ratio when $\phi_{\rm{NR}} \lessapprox 0.5$. 


\section{Further discussions}

\subsection{Granular Temperatures}

\begin{figure}
	\centering
	\includegraphics[scale = 0.45]{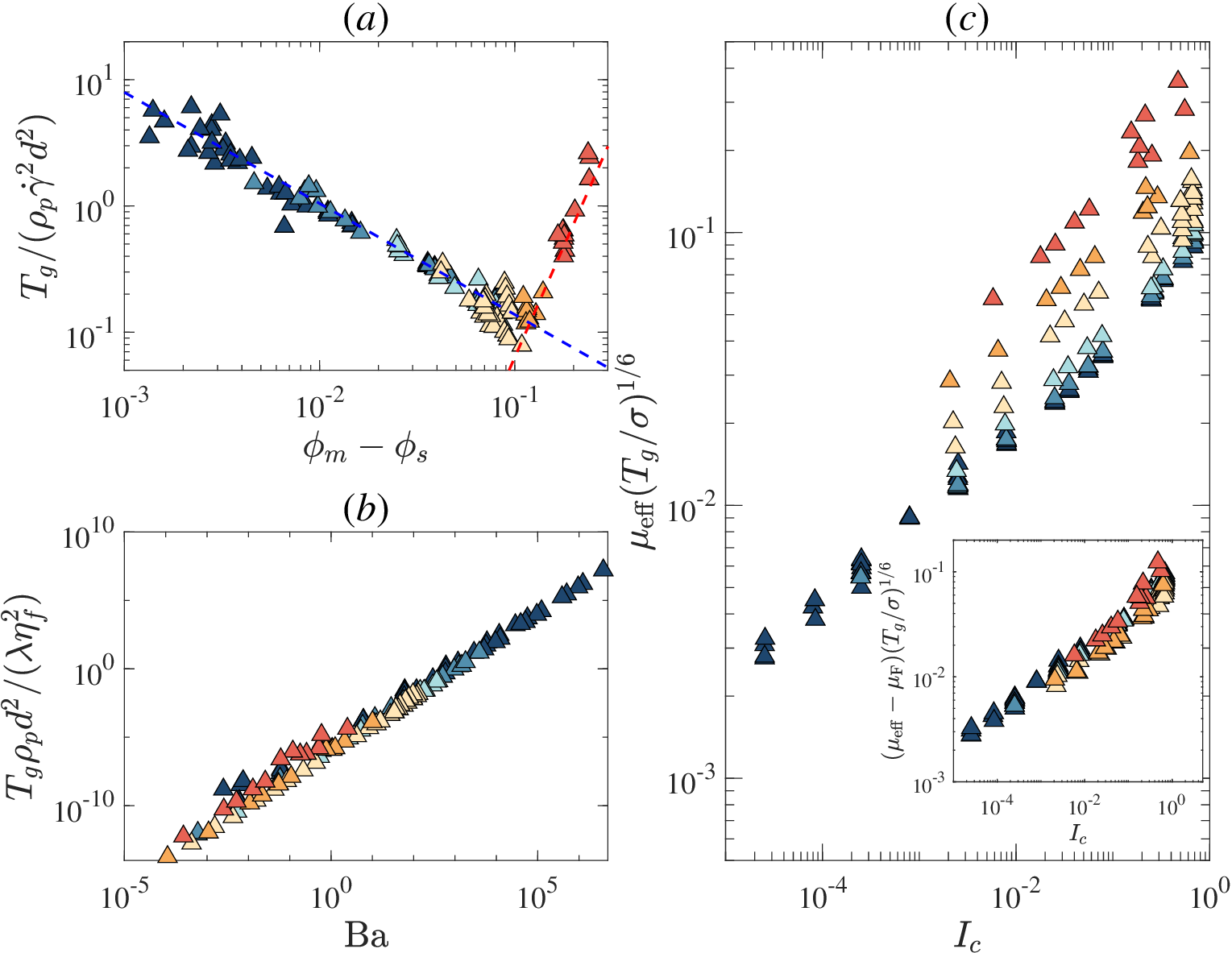}
	\caption{(a) Relationship between normalized granular temperature, $T_g/(\rho_p\dot{\gamma}^2d^2)$ and $\phi_m-\phi_s$ using a log-log scale. (b) Relationship between $T_g\rho_p d^2/(\lambda\eta_f^2)$ and Bagnold number $\rm{Ba} = \lambda^{0.5}\rho_p d^2\dot{\gamma}/\eta_f$. (c) We plot our simulation results as the relationship between $\mu_{\rm{eff}}(T_g/\sigma)^{1/6}$ and $I_c$, which was proposed by \citet{kim2020}. The markers in this figure are the same as those in Fig. \ref{fig:12}(b).}
\label{fig:14}
\end{figure}

To better understand the behaviour of granular-fluid systems, we also investigate the \textit{granular temperature} given different pairs of $(I_c, I_v)$. We define the granular temperature of the sheared granular-fluid system as $T_g \equiv \rho_p\langle (\vec{u} - \langle \vec{u} \rangle)^2 \rangle$, i.e., $T_g$ has the same dimension as stresses, and plot its relationship with $\phi_m-\phi_s$ in Fig. \ref{fig:14}(a). Quite unexpectedly for us, the granular temperature $T_g$ plotted vs.\ $\phi_m-\phi_s$ does not help us to distinguish between collisional and visco-collisional flows. Rather, for nearly all of our systems, $T_g$ has the same functional dependence on $\phi_m-\phi_s$, not dissimilar from that predicted by Bagnold for moderate-to-dense granular-fluid flows \citep{bagnold1954} shown in Figs. \ref{fig:14}(b). The exceptions we found (not surprisingly) were for $\phi_s \lessapprox 0.51$ and $I_v>0.1$ for which we found that $T_g$ increases with increasing $\phi_m-\phi_s$ (Fig.\ \ref{fig:14}(a)). In Fig. \ref{fig:14}, we suggest that the normalized granular temperature $T_g/(\rho_p\dot{\gamma}^{2}d^{2})$ can be fitted with $T_g/(\rho_p\dot{\gamma}^{2}d^{2}) = (1/55)(\phi_m-\phi_s)^{-0.88}$ when $\phi_s\gtrapprox0.5$ (blue dashed line in Fig. \ref{fig:14}(a)), but fitted with $T_g/(\rho_p\dot{\gamma}^{2}d^{2}) = 200(\phi_m-\phi_s)^{3.5}$ when $\phi_s\lessapprox0.5$ (red dashed line in Fig. \ref{fig:14}(a)).

We compare these results with the predictions of \citet{bagnold1954} according to his theory relating stresses to rates of average interparticle collisions, distances, etc., based on imposed conditions. With a relatively simple collisional model, Bagnold derived a form for a normalized dispersive stress that scales with $\lambda \equiv 1/[(\phi_0/\phi)^{1/3} - 1]$.  His data suggest that a normalized form of $\tau_c$ scales with $\lambda^2$ when the effects of grain inertia dominate, and that the normalized dispersive stress scales with $\lambda^{1.5}$ when the effects of fluid viscosity dominate. Here, $\phi_0$ is the maximum possible static volume fraction of granular material and is equal to 0.74. In Fig. \ref{fig:14}(b), we plot the relationship between $T_g\rho_p d^2/(\lambda\eta_f^2)$ and Bagnold number $\rm{Ba} = \lambda^{0.5}\rho_p d^2\dot{\gamma}/\eta_f$. Similarly to the results obtained by Bagnold, the normalized granular temperature (kinetic pressure) scales roughly with the squared Bagnold number, except that the data are more scattered when Ba is small, which corresponds to the red triangles shown in Fig. \ref{fig:14}(a).

In a previous study, \citet{kim2020} proposed that the multiplication of $\mu_{\rm{eff}}$ and a dimensionless granular temperature, $\Theta^{1/6}$, where $\Theta = T_g/\sigma$, could yield a monotonously increasing scaling with respect to the inertial number. We note that, since our $T_g$ already incorporates the particle density in its definition, $T_g$ naturally has the same dimension as stresses. Thus, we plot the relationship between $\mu_{\rm{eff}}(T_g/\sigma)^{1/6}$ and $I_c$ in Fig. \ref{fig:14}(c), which shows that, when $I_v<0.001$, our simulation data collapse onto a master curve similar to what was proposed by \citet{kim2020}. However, as soon as we increase $I_v$ that the the system transitions to a visco-collisional regime $\mu_{\rm{eff}}(T_g/\sigma)^{1/6}$ starts to deviate from the master curve, which implies that the influence of $I_v$ should also be incorporated. Looking back at our analyses in Sect. \ref{sec:fric_rheo}, it may be possible to substract the influence of the viscosity induced frictional rheology from the effective frictional coefficient. Thus, we also calculate $\mu_{\rm{eff}} - \mu_{\rm{F}}$, where $\mu_{\rm{F}} = \mu_{\rm{3v}}/(1+I_{\rm{v0}}/I_v)$, $\mu_{\rm{3v}} = 1.1$, and $I_{\rm{v0}} = 0.05$ are what we have obtained from Eqn. \ref{eq:mueff1}, and plot the relationship between $(\mu_{\rm{eff}} - \mu_{\rm{F}})(T_g/\sigma)^{1/6}$ and $I_c$ in the inset of Fig. \ref{fig:14}(c). After eliminating the influence of the interstitial fluid, the scaling of $(\mu_{\rm{eff}} - \mu_{\rm{F}})(T_g/\sigma)^{1/6}$ and $I_c$ approximately collapses onto a master curve, as suggested by \citet{kim2020}. This also showcases that our partition of $\mu_{\rm{eff}}$ based on $\mu_{\rm{inert}}$ and $\mu_{\rm{F}}$ works well for granular-fluid systems.

Comparison between our data and previous research shows that our data for granular temperature have similar scaling as those of dispersive stresses in \citet{bagnold1954} and kinetic stresses in \citet{lun1991}, except when the solid fraction is so large that the system is approaching a jamming state. As what we have previously stated, a more sophisticated investigation, such as coupled LBM-DEM simulations, is still needed to explore beyond the $\mu-I_c$, $\mu-I_v$ or $\mu_{\rm{eff}}(\mathcal{A}G^{0.5})$ relationship to elaborate both the hydrodynamic effect and the collisional effect to obtain a better stress partition within granular-fluid systems.

\subsection{Lubrication effect and drag forces}

\begin{figure}
	\centering
	\includegraphics[scale = 0.45]{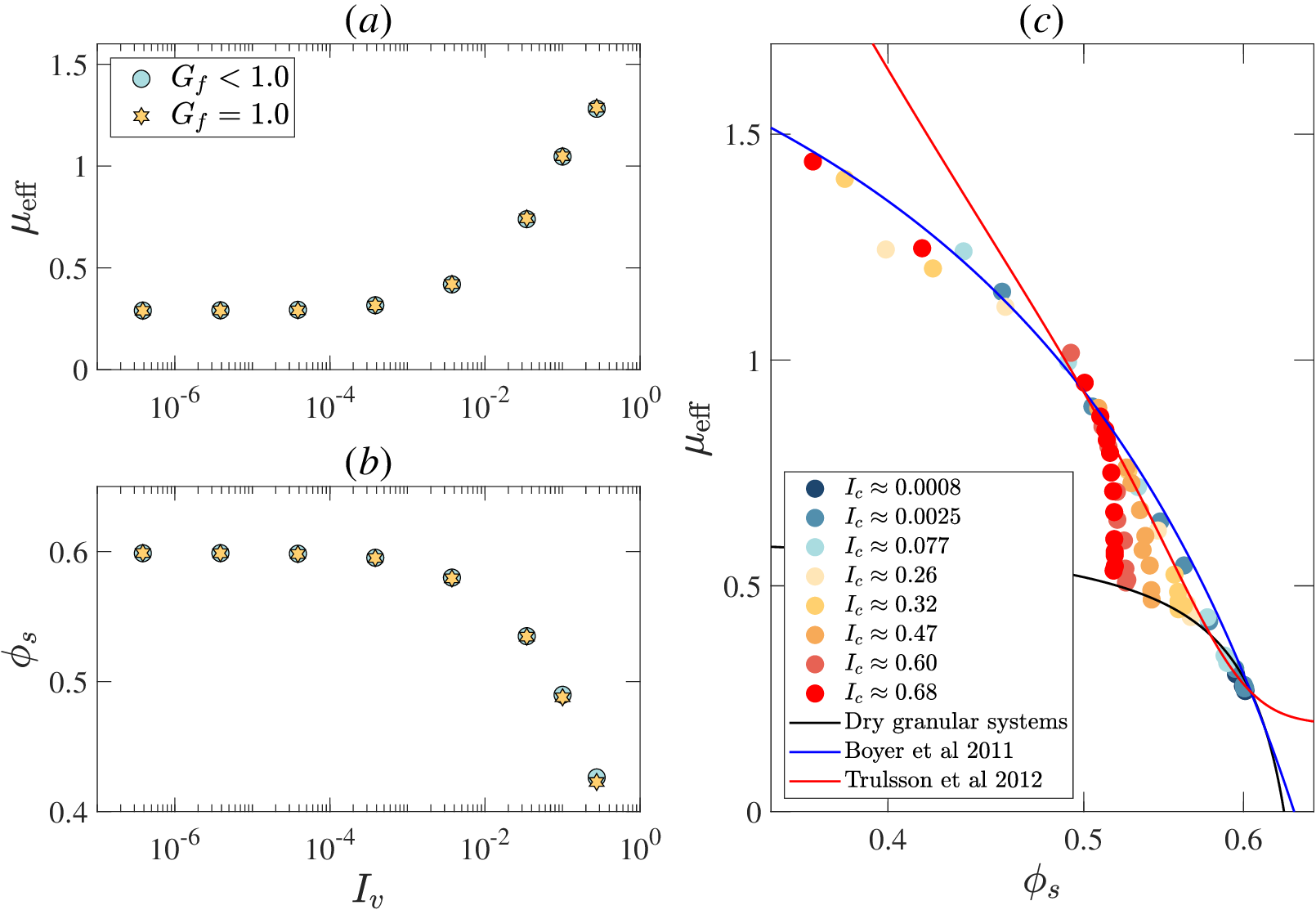}
	\caption{(a) Comparison between two different ways of calculating the lubrication effect with the $\mu_{\rm{eff}}-I_v$ relationship. In these two sets of simulations, no drag force was included. (b) Comparison between two different ways of calculating the lubrication effect with the $\phi_s-I_v$ relationship. For simulations within Figures (a) and (b), the inertia number, $I_c$, is approximately 0.025. (c) Simulation results of DEM with both lubrication and drag forces plotted as the relationship between $\mu_{\rm{eff}}$ and $\phi_s$. Dashed lines are the same as those presented in Fig. \ref{fig:9}.}
\label{fig:13}
\end{figure}

In previous sections, we have addressed the significance of the interstitial fluid, but this influence is built on the numerical approach we implement. In the simulation, to calculate the effect of interstitial fluid on adjacent particles, we regard the lubrication effect as an effective coating on the surface of particles, i.e., the spherical lubrication zone. Thus, when we calculate the lubrication force, a reduction factor, $G_f$ (shown in Eqn. \ref{Gf}), needs to be considered. Also, as mentioned previously, we use two regularizing length scales for this viscous force: (1) $\delta_{\rm{g,min}}\sim \langle d \rangle /10 \approx 2t$ can be thought of as twice a typical particle roughness length scale, $t$, and (2) $\delta_{\rm{g,max}} \sim \langle d \rangle \approx 2H$ (twice the maximum lubrication length scale, $H$). Thus, $G_f$ varies naturally from $\approx$0.9 to $\approx$0.2 as shown in Fig. \ref{fig:1}(b), which is different from other works \citep{trulsson2012} where $G_f$ is constant at $1.0$. To clarify the influence of $G_f$, we also simulated the simple shear test with $G_f = 1.0$. The results are shown in Figs. \ref{fig:13}(a) and (b), which show that the effect of using $G_f < 1$ in the range relevant to our problem is essentially negligible and the reduction of lubrication area does not results in a significant change in the macroscopic behaviour of granular-fluid systems.


Further, we also investigate the importance of drag forces in a sheared granular-fluid system. In a sheared granular system, once it reaches the steady-state, a linear velocity profile is expected for both the solid particle phase and the fluid phase. In the previous section, we assume that the velocity differences between particles and fluids are negligible, so that we deliberately neglect the drag effect to have a simple DEM-lubrication model framework. In this section, because the velocity difference between particles and surrounding fluids is small and the granular-fluid system is mainly within a low-Reynolds number state, we add a Stoke's drag to each grain so that
\begin{equation}
    \begin{split}
        F_{\rm{d,k}}^{\rm i} = 3\pi d_p\eta_f(v_{\rm{f,k}} - v_{\rm{p,k}})\ , \label{eq:drag}
    \end{split}
\end{equation}
where $F_{\rm{d,k}}^{\rm i}$ denotes the drag force in the $k-$direction subjected to particle $i$ and $k = X, Y, $or $Z$, $v_{\rm{f,k}}$ is the average $k-$direction fluid velocity in the neighbourhood of particle $i$, and $v_{\rm{p,k}}$ is the grain velocity in $k-$direction. We assume that the fluid velocity has a perfectly linear profile along $X-$direction, i.e., the height direction, and the fluid velocities in both $Y-$ and $Z-$ directions are zero. In this set of simulations, we vary the normal pressure from 100 Pa to 1000 Pa, the interstitial fluid viscosity from 0 cP to 10$^{4}$ cP (when $\eta_f\neq 0$, the minimum fluid viscosity is in the order of 10$^{-4}$ cP), and the shear rate from 0.1 s$^{-1}$ to approximately 46 s$^{-1}$.

Figure \ref{fig:13}(c) shows the results obtained from simulations with drag forces, where we plot the relationship between $\mu_{\rm{eff}}$ and $\phi_s$, and the solid lines are the same as those presented in Fig. \ref{fig:9}. We note that, in this set of simulations, we regenerate the granular assembly so that their particle size distribution, while it keeps the same mean and variation, may be slightly different from the previous set of simulations. Fig. \ref{fig:13}(c) shows that all the simulation data still fall within the classical $\mu_{\rm{eff}}(I_c)-\phi_s(I_c)$ curve for dry granular systems and the visco-inertial $\mu_{\rm{eff}}(I_s)-\phi_s(I_s)$ curve proposed by \citet{trulsson2012}. When the fluid viscosity is zero, the $\mu_{\rm{eff}}-\phi_s$ relationship more or less follows the dry granular $\mu_{\rm{eff}}(I_c)-\phi_s(I_c)$ curve, which is the same as that shown in Fig. \ref{fig:9}.

When the viscosity of interstitial fluid is relatively large, the effective frictional coefficient, $\mu_{\rm{eff}}$, follows approximately the fitting curve proposed by \citet{trulsson2012}. When $\phi_s<0.45$, $\mu_{\rm{eff}}$ obtained from this set of simulations is slightly smaller than those in Fig. \ref{fig:9}. However, this does not contradict our previous conclusion that the relationship between $\mu_{\rm{eff}}$ and $\phi_s$ is complex and almost impossible to have a one-on-one monotonic relationship. We also note that the partition of hydrodynamics forces into drag forces and lubrication forces is still questionable. The summation of drag forces and lubrication forces may exceed the real hydrodynamic effect when particles are positioned inside a fluid. Thus, it will be more reasonable to fully resolve the particle-fluid interaction using numerical tools such as  the coupling method of lattice Boltzmann method (LBM) and discrete element method. This further study on the partition of hydrodynamic forces will be investigated in future works.

\subsection{Extension of Simulation and Theory to other Frameworks}
\label{sec:eg}
\begin{figure}
  \centering
  \includegraphics[scale = 0.5]{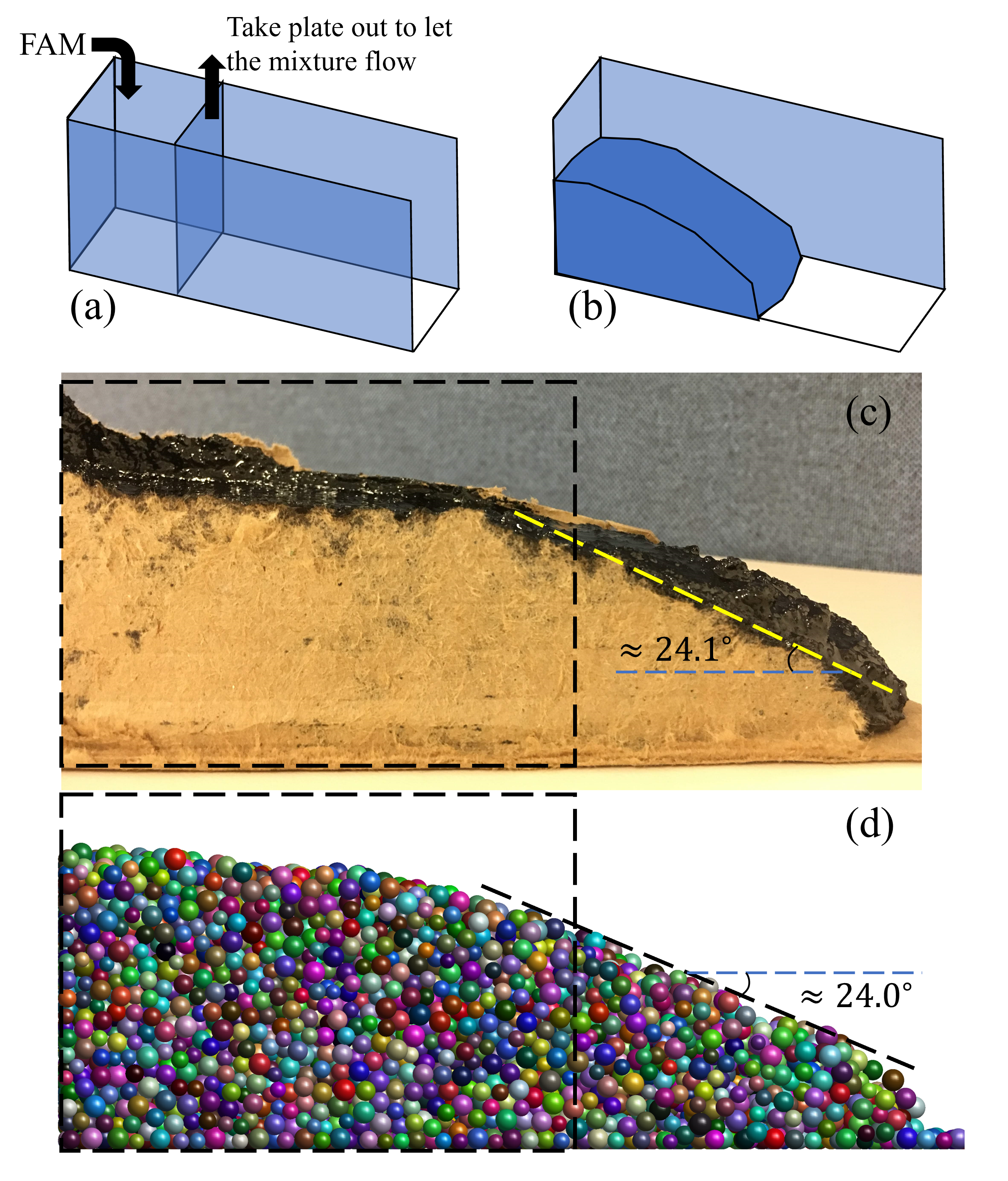}
  \caption{A physical example and its corresponding DEM simulation. (a) at the initial state when we have heated the testing channel to a certain temperature, we put FAM into the designed space to form a packing of saturated granular-fluid system, after which we could take the plate out to let the mixture flow. (b) shows the sketch of the final state of the granular column collapse. (c) the final state of the experiment of the FAM with a repose angle of approximately 24.1$^{\circ}$. (d) the final state of the DEM simulation with lubrication forces, which shows a repose angle of approximately 24$^{\circ}$.}
  \label{fig:2}
\end{figure}

To validate the model framework, we simulate the collapse of a granular column (a fully saturated mixture of asphalt binder and sand, namely fine aggregate mastic or FAM)  using the same DEM model as we have described in the previous section and compare the simulation result with its experimental counterpart. The experiment and the simulation are designed as shown in Fig. \ref{fig:2}(a,b). In the physical test of the FAM column collapse, we use the asphalt binder of PG58-28, where ``PG" stands for performance graded design, and ``58-28" represents the service temperature of the asphalt binder and corresponds to its service properties and is out of scope of this work, since we, in this paper, are only concerned with its viscosity at certain temperature when the asphalt binder is in its liquid phase. We tested that, between \SI{100}{\celsius} and \SI{150}{\celsius}, the viscosity of the PG58-28 asphalt binder varies from 2000 cP to 120 cP. Since our experiment is performed at \SI{150}{\celsius}, the viscosity of the interstitial fluid in the simulation is set to 120 cP. The fine aggregates are river sand, with a median size of $d_{50}\approx 1.0$ mm. In the simulation, we set the mean particle size to 1.0 mm, but impose a uniform grain size distribution (GSD) varying from 0.5 to 1.5 mm. Since the GSDs of the experiment and the simulation are different from each other, we expect slight differences in their collapse behaviours, but the final run-out distance and the deposition pattern should be similar.


In the experiment, we first heat the testing channel and FAM to the testing temperature (\SI{150}{\celsius}) in the oven. After half an hour, we remove both the channel and the FAM from the oven, and place the FAM into the confining area of the channel, then put the whole device back into the oven for 5 minutes to reheat the material to \SI{150}{\celsius}. Then we take the device out, remove the confining plate, and let the FAM flow to the opening end of the channel. The FAM flow stopped quickly; no obvious temperature change was observed on the surface of the flow measured using a hand-held thermometer. After one day of the test, the FAM had become solid state. We remove the cardboard channel, so that we can clearly see the interface line (shown in Fig. \ref{fig:2}) between the channel and the FAM. Using this interfacing line, we can define the angle of repose of the FAM at certain temperature. The shape of the FAM after tests will be compared with the DEM simulations.

We simulate the collapse of the granular column using the same DEM model as we previously described. The viscosity of the interstitial fluid, i.e. asphalt binder, in the simulation is the same as that in the physical test. In the simulations, the elastic modulus of the particle is $E = 29$ GPa, the Poisson ratio is $\nu = 0.2$, and the particle density is $\rho_p =$ 2650 kg/m$^3$. We also set the inter-particle frictional coefficient to be 0.3 and the restitution coefficient to be 0.2. Fig. \ref{fig:2} shows the comparisons between the experimental result (at \SI{150}{\celsius}, which is the mixing temperature for the hot mixed asphalt) and the simulation result (with viscosity the same as the experimental results in the binder viscosity test, 120 cP). As we can see from the comparison results, the simulations can reproduce the experimental result well in terms of both the run-out distance and the deposition angle. Consequently, we continue to use this DEM-lubrication framework to study further the rheological behaviour of granular-fluid systems. Furthermore, in our previous papers, we utilized the same DEM-lubrication model to construct a two-scale simulation framework to calculate the gyratory compaction behaviour of hot-mixed asphalt, which exhibits good consistency with the experimental results \citep{man2021granular,man2022two}.

\section{Conclusions}
\label{sec:conclud}

To summarize, using discrete element simulations we investigate the rheological behaviour of granular-fluid flows, where the effect of the interstitial fluid is considered as a spherical lubrication zone. We demonstrate that our simulation results acquired from this simple contact-lubrication model show the same behaviour as suggested theoretically by \citet{Ge2023unifying}, which was validated with experimental results in \citet{midi2004,boyer2011,Tapia2022}. With such a simple model, we are able to obtain the constitutional behaviour of granular-fluid systems from two aspects:
\begin{itemize}
    \item We decompose the effective frictional coefficient, $\mu_{\rm{eff}}$, and the average solid fraction, $\phi_s$, into two parts. In so doing, we write the frictional rheology and dilatancy law of granular-fluid systems as combined functions of both the inertial number $I_c$ and the viscous number $I_v$, which shows a good collapse of all the simulation data.
    \item On the particle scale, we decompose each particle interaction into the collisional part and the lubrication part, which lead to the calculation of both the collisional shear stress due to particle collisions and the viscous shear stresses due to the interstitial fluid. Based on this, we can construct the combined equation for $\mu_{\rm{eff}}$, which is a function of $(\phi_s, I_v, I_c)$.
\end{itemize}

Thus, we could attribute the rheological behaviour of granular-fluid flows to both effects from $I_c$ and $I_v$. However, simultaneous variation of both effects reveals concentration-dependent transitions in bulk behaviours reflected in interparticle forces and force networks not previously reported for similar particle-fluid systems. Based on the simulation results, there exists no one-on-one monotonic relationship between $\mu_{\rm{eff}}$ and $\phi_s$, which indicates that the combination of the collisional effects and the viscous effects is nonlinear.

This richness is reflected even in relatively simple fabric measures such as the relationship between coordination number and ``non-rattler'' particle fraction (fraction of particles with a minimum of two contacts \citep{bi2011jamming}).  Links like this give insight to a more broadly unified physical framework for dense particle-fluid deformation and flows. In exploring this in detail, we provide physics-based intuition on the physics of the phase transitions between collisional, viscous, and visco-collisional granular flows. This article finishes with additional observations on the granular temperature to show that the scaling of granular temperatures presented in this work is similar to experimental works of \citet{bagnold1954}, and the incorporation of granular temperatures helps unify the scaling of the $(\mu_{\rm{eff}}-\mu_{\rm{F}})(T_g/\sigma)^{1/6}\sim I_c$ relationship, which is similar to that presented in \citet{kim2020}. Meanwhile, we note that detailed investigations are still needed to fully understand the particle-fluid interactions in dense sheared granular-fluid flows. Thus, a thorough and fully resolved LBM-DEM coupled simulation is preferred in future studies. Ongoing work also includes investigations of the transitions between particle-scale interactions of these systems and dynamics in analogous fully saturated systems.

\begin{acknowledgements}
\textbf{Acknowledgements}- We gratefully acknowledge the funding for this research provided by the NSFC under grant No. 12202367 and by the NSF under the grant EAR-1451957. We also thank the support from the UMN Center of Transportation Studies, the CEGE Sommerfeld Fellowship, and computing resources provided by Saint Anthony Fall Laboratory at UMN and by High-Performance Computing Center at Westlake University. The authors also thank Prof.\ J.-L. Le, Prof. S. A. Galindo-Torres, and Dr. Z. Ge for helpful discussions. K.M. Hill would like to thank Prof. \'E. Guazzelli for discussions during the Gordon Research Conference and other occasions.

\textbf{Declaration of Interests}- The authors report no conflict of interest.
\end{acknowledgements}


\bibliographystyle{jfm}
\bibliography{visco-collision}

\end{document}